\documentclass{eas}

\usepackage{graphicx}
\usepackage{lscape}
\usepackage{natbib}
\usepackage{txfonts}

\newcommand{\arcsec}{''}
\newcommand{\degr}{^{o}}
\def\aj{AJ}
\def\apj{ApJ}
\def\aap{A\&A}
\def\aaps{A\&AS}
\def\mnras{MNRAS}
\def\apjl{ApJL}
\def\nat{Nature}
\def\araa{Annual Rev. of Astron. \& Astrophys.} 

\TitreGlobal{What can the highest angular resolution bring to stellar astrophysics?}

\begin{document}
\title{Observation of Double Star by Long Baseline Interferometry}
\runningtitle{double star interferometry}

\newcommand{\folder}{./}

%
\author{D. Bonneau} \author{F. Millour} \author{A. Meilland}
\address{Laboratoire Lagrange, UMR7293, Universit\'e de Nice Sophia-Antipolis, CNRS, Observatoire de la C\^ote d'Azur, Bd. de l'Observatoire, 06304 Nice, France\\\email{fmillour@oca.eu}}

\begin{abstract}
 This paper serves as a reference on how to estimate the parameters
 of binary stars and how to combine multiple techniques, namely
  astrometry, interferometry and radial velocities.
\end{abstract}



\maketitle
%
\section{Why binary stars matter?}
\label{Sect:intro}

The concept of physic stellar couples was imposed by astrometric
observations of visual binaries in the XIXth century and the
spectroscopic and photometric binaries in the early XXth century \citep{Heintz1978}.  
It quickly became apparent that the most interesting visual double stars
were the one with a fast orbital motion (throughout an astronomer's
life ...) so that with a small angular separation between the
components whose astrometric measurement required high angular
resolution (i.e. less than 1 arcsecond). 
It was found that the star formation process leads to extremely frequent 
stellar multiplicity with about two thirds of the stars belonging to a 
multiple system \citep{Duchene2013}.
It was also found that the components of double stars could belong 
to all spectral types (related to temperatures and stellar masses) 
and that the physical separation of the components could vary considerably 
between wide binaries with separations large compared to the size of 
the stars (usually seen as visual binary or long period spectroscopic binary) 
and close binaries (observed as spectroscopic and photometric binaries) 
with separations which can be of the same order of magnitude as the size of the
stars \citep{guinan2007}. 
In this case, phenomena occur relating to the interaction between the components 
(mass transfer, circumstellar or circumbinary accretion disks, jets...).

When modeling stars, astrophysicists make very often the assumption
that they were born, evolve and will die as a single star. This
assumption, probably acceptable for wide binaries, is thrown into
doubt more and more often for many types of stars and is no longer
usable for close binaries. Indeed, for close binaries, the
interactions that occur between components affect their evolutions no
more similar to those of single stars.  For example, very massive
stars in the main sequence (O-type stars) were thought to be single a
decade ago, but thanks to recent works of \cite{Sana2012}, it has been
demonstrated that almost all O stars are multiple. At the same time,
recent developments in evolutionary models of stars allow one to
consider the presence of a companion star (see for
example \citep{Siess2013}) and indeed it changes significantly the
evolutionary tracks of stars.

The input of high angular resolution observations is fundamental to
the study of the morphology of binary systems in addition to data
provided by the spectroscopic and photometric observations.  That is
why the observation of double stars has become one of the main fields
of application of interferometric techniques developed for high
angular resolution astronomical observations. For about forty years,
visual double stars and some spectroscopic binaries are accurately
measured by speckle interferometry \citep{Hartkopf2001}, but in the
last twenty years, the measurement and morphological study of the
closer systems became possible by long baseline interferometry (see
for example \citet{McAlister2007} and \citet{Davis2007}).  However we
do not forget that, to be fully usable, interferometric observations
should be combined with spectroscopic, photometric, astrometric
(Hipparcos, Gaia) observations, and the use of physical models of
objects adapted to provide interferometric observable predictions
(i.e. amplitude and phase of the fringe visibility).

Thus, the interferometric observations of double stars contribute
\begin{itemize}
\item To the determination of the fundamental stellar parameters (masses, radii); 
\item To the study of stellar evolution, from the initial phases (star formation, Initial Mass Function) at the end of life of the stars.
\end{itemize}


\section[Fundamental parameters]{The Study of binary stars to determine stellar fundamental parameters}
\label{fundpara}

Validation of knowledge in the field of stellar physics relies heavily
on our ability to confront the physical parameters (such as
temperature, mass, and radius) predicted by the model and obtained
from observations.  We will see in this section that the determination
of these parameters is possible by combining the interferometric
observations with spectroscopic, astrometric, and photometric
observations, depending on the class of the observed binary.

\subsection{Components of a double line spectroscopic binary resolved as a visual binary}
\label{subSect:fundparaSB2}

\begin{figure}[htbp]
\centering
\includegraphics[width=0.9\textwidth]{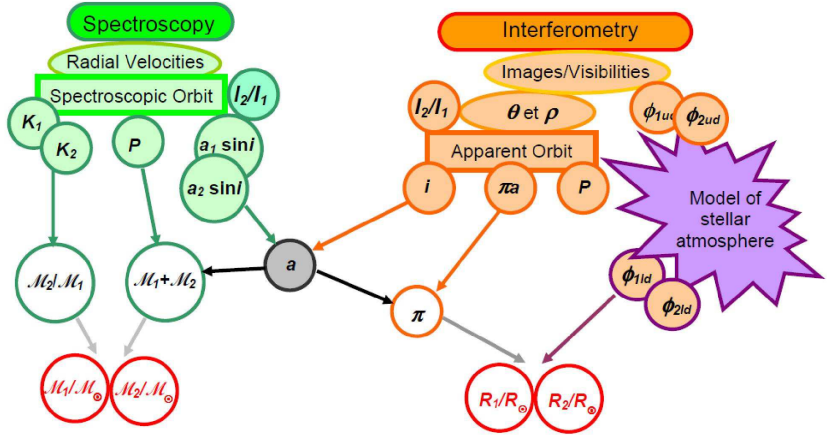}
\caption[]{
Determination of Masses and Radii for a resolved double line spectroscopic binary}
\label{Fig:fundparaSB2}
\end{figure}

For a double line spectroscopic binary (SB2), the radial velocity
measurements provide the spectroscopic orbits of the two components
around the center of mass of the binary system.  The case of a SB2
binary resolved as a visual double star whose stellar discs are
resolved by interferometry is most favorable for the determination of
stellar fundamental parameters (see Figure~\ref{Fig:fundparaSB2}).

It should first be noted that the distance of the system can be
determined by combining the spectroscopic and apparent orbits.  The
accuracy of the orbital parallax depends more on the quality of
spectroscopic and interferometric measurements than of the distance of
the system.  Then, the values of the stellar masses are then directly
obtained by combining the visible and spectroscopic orbits.

Moreover, the use of a model of stellar atmosphere is necessary to
obtain the value limb-darkened angular diameter $\phi_{ld}$ from the
uniform disc angular diameter $\phi_{ud}$ given by interferometric
measurements.

The value of the intensity ratio of the two components $I_2/I_1$ can
be derived from spectroscopic analysis as well as the visibility
measurements (see \ref{subSect:doublesource}).  Further
spectrophotometric data are needed to estimate the effective
temperature and the luminosity of each system components.

\subsection{Components of a single line spectroscopic binary resolved as a visual binary}
\label{subSect:fundparaSB1}

\begin{figure}[htbp]
\centering
\includegraphics[width=0.9\textwidth]{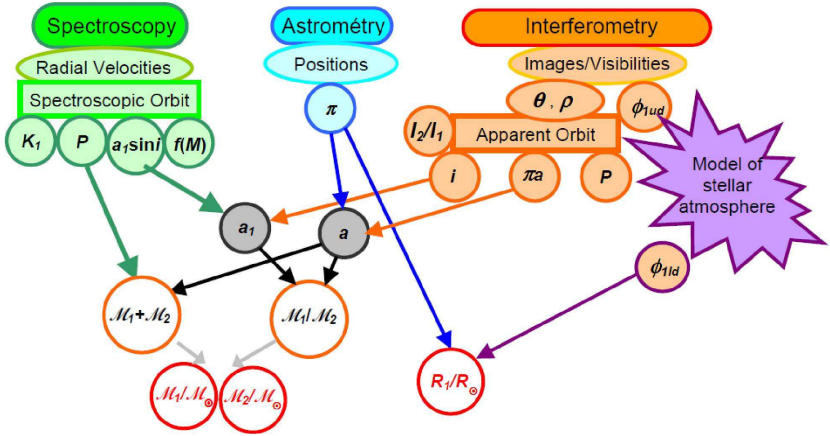}
\caption[]{
Determination of Masses and Radii for a resolved single line
spectroscopic binary}
\label{Fig:fundparaSB1}
\end{figure}

For a single line spectroscopic binary (SB1), the radial velocity
measurements provide only the spectroscopic orbit of the brightest
component.

Interferometric observations resolve the system as a visual binary and
allow computing the apparent orbit and giving an estimate the
intensity ratio of the components (see Figure~\ref{Fig:fundparaSB1}).

Generally, only the angular diameter of the primary component can be
estimated from the interferometric visibility analysis. The use of a
model of stellar atmosphere is necessary to obtain the value
limb-darkened angular diameter $\phi_{ld}$ from the uniform disc
angular diameter $\phi_{ud}$.  The determination of the stellar masses
and radii requires knowledge of the distance of the system determined
from astrometric observations.  Further spectrophotometric data are
needed to estimate the effective temperature and the luminosity of
each system component.

\section[Orbit]{The orbit of a binary star}
\label{Sect:binaryorbit}
There is an extensive literature on the subject of double stars and
many websites are devoted to this topic and much information is
available on the Web site of the "{\it{Double Star
Library}}"~\footnote{http://www.usno.navy.mil/USNO/astrometry/optical-IR-prod/wds/dsl}.
We only recall some basic and the contribution of interferometric
observations to the measurement of double stars.

\begin{figure}[htbp]
\centering
\includegraphics[width=0.4\textwidth]{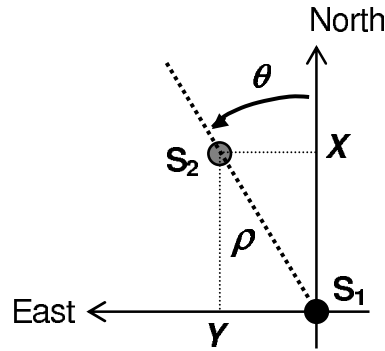}
\caption[]{
Measure of a {\it{visual double star}}}
\label{Fig:visualbin}
\end{figure}

The measure of a binary star consists in recording the relative
positions of the components in a tangent plane to the celestial sphere
(the sky plane) at a given epoch $t$.  As shown in
Figure~\ref{Fig:visualbin} the coordinates of the B component relative
to the position of the A component are:

\begin{equation}
\label{relativeposi}
X = \rho~cos \theta = \Delta \alpha~cos \delta, \quad  		
Y = \rho~sin \theta = \Delta \delta.
\end{equation}

With, the position angle $\theta$ (in degree) measured from the North
($0^\circ$) to the East ($90^\circ$) and the angular separation $\rho$
(in arcsecond).

The main purpose of the double star observations is to obtain orbits
whose knowledge is the key to determining the masses and even stellar
distances, fundamental parameters of stellar astrophysical studies.
The contribution of the interferometric observations in addition to
spectroscopic and photometric observations is valuable for the
astrophysical study of double stars.

The orbit of a double star reflects the motion of the stars under the
influence of gravity and therefore obeys the Newton's and Kepler's
laws.

\begin{figure}[htbp]
\centering
\includegraphics[width=0.7\textwidth]{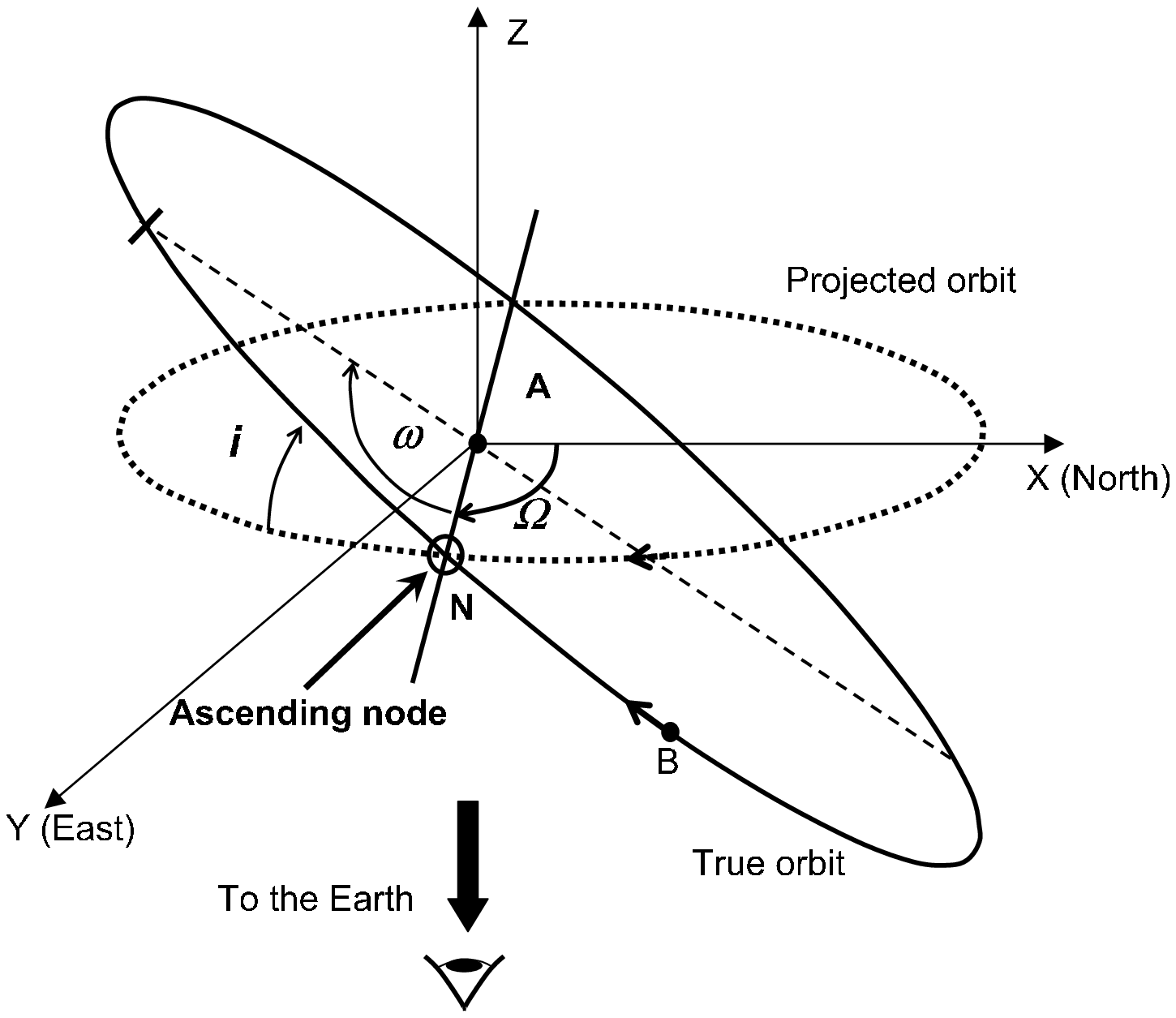}
\caption[]{
The true and the projected (B/A) orbits of a {\it{visual double star}}
and its geometrical elements}
\label{Fig:orbitelements}
\end{figure}

Three types of orbits may be considered to describe the orbit of a
double star (components A and B) bound by the gravitation:
\begin{itemize} 
\item The (A/AB) and (B/AB) {\it{absolute orbits}} of A and B
      around the center of mass AB respectively.  For resolved
      binaries, astrometric observations give then access to the
      (AB/A) and (AB/B) orbits projected on the sky plane and
      spectroscopic measurements of the radial velocity allow to
      determine the (AB/A) and (AB/B) orbits projected on the line of
      sight. In the case of a binary with large magnitude difference,
      the astrometric measurements capture the position of the
      photocenter F located between A and AB which determines the
      (F/AB) {\it{astrometric orbit}} with a semi-major axis aF and
      radial velocity measurements give only access to the (A/AB)
      orbits projected on the line of sight.
\item The {\it{relative orbit}} (B/A) of B around A
      (Figure~\ref{Fig:orbitelements}).  The (B/A) orbit projected on
      the sky plane, known as the {\it{apparent orbit}}, can be
      derived from measurements of the relative position of the
      components obtained with astrometric imaging or interferometric
      observations.
\end{itemize}

It should be noted that, for a given projected orbit on the sky plane,
there is two possible true orbits, which are symmetric with respect to
the sky plane.

\subsection{Classical orbital parameters}
\label{subSect:orbitalelements}

The spatial motion of the binary components is described using the
reference frame (A, x, y, z) centred on the A component with two axes
in the plane tangent to the celestial sphere {\it{plane of the sky}}:
Ax axis points towards the North (position angle = $0^\circ$), Ay axis
points towards the East (position angle = $90^\circ$). The Az axis is
along the line of sight, pointing in the direction of increasing
radial velocities (positive radial velocity).

This reference frame is thus retrograde: viewed from the positive side
of the Az-axis, a rotation of the Ax-axis onto the Ay-axis is carried
out clockwise.  The relative (B/A) orbit is described by means of
seven {\it{classical orbital elements}}.

Four so-called {\it{dynamic elements}} specifying the properties of
the Keplerian motion in the true orbit:
\begin{itemize}
\item $T$, the epoch of passage through periastron expressed in Julian days or in Besselian years \footnote{the "Besselian epoch" can be calculated by $B = 1900.0 + (Julian~date - 2415020.31352) / 365.242198781$}. 
\item $P$, the revolution period (in day or in year)
\item $a$, the true semi-major axis (in km or au\footnote{astronomical unit is written "au", instead of "AU", since
the 2012 IAU resolution B2 "on the re-definition of the astronomical
unit of length"}) or $a\arcsec = \varpi a$, the apparent semi-major axis (in arcsecond)
\item $e$, the eccentricity
\end{itemize}

Three so-called {\it{geometric elements}} which define the orientation of the orbits (see Fig.~\ref{Fig:orbitelements}) are:
\begin{itemize}
\item $\Omega$, is the position angle of the line of intersection between
      the true orbital plane and the plane of the sky. There are two
      nodes whose position angles differ by $180^\circ$but by
      convention, $\Omega$ is the position angle of the {\it{ascending
      node}} (where the orbital motion is directed away from the
      Sun). If radial-velocity measurements are not available to
      identify the {\it{ascending node}}, a temporary value ranging
      between $0^\circ$and $180^\circ$ is adopted following a
      convention proposed by~\citep{Finsen1934}.
\item $i$, the inclination, is the angle between the planes of the
      projected orbit and of the true orbit, taken at the ascending
      node. The motion is prograde if $0^\circ < i < 90^\circ$ and
      retrograde if $90^\circ < i < 180^\circ$.
\item $\omega$, is the argument of the periastron in the true orbit plane,
      measured in the direction of the orbital motion from the
      {\it{ascending node}} with a value ranging from $0^\circ$ to
      $360^\circ$.
\end{itemize}

\subsection{Total mass of the system}
\label{subSect:totalmass}
For a binary system whose relative (B / A) orbit and the distance is
known, the total mass can be calculated by application of Kepler's
third law.
\begin{equation}
\label{totalmass}
M = M_A + M_B = \frac{a^3}{\varpi^3~P^2},
\end{equation}

with, $\varpi = 1/d$, the parallax (in arcsecond) and $d$ the distance
(in parsec) of the double star, $M$ the total mass of the system in
solar mass $M_{\odot}$, $a$ the semi-major axis (in au), and $P$ the
orbital period (in year).

The relative error on the total mass can be estimate with:

\begin{equation}
\label{errormass}
\frac{\Delta M}{M} \approx 3~\frac{\Delta a}{a} + 2~\frac{\Delta P}{P} + 3~\frac{\Delta \varpi}{\varpi}.
\end{equation}

It should be noted that for a well determined orbit, errors on
parameters $a$ and $P$ are independent and the error on the mass of
the system is dominated by the parallax error.
 
\subsection{The true orbit}
\label{subSect:trueorbit}
The true (B/A) orbit is the Keplerian elliptical orbit of the B around A.

\begin{figure}[htbp]
\centering
\includegraphics[width=0.7\textwidth, angle=0]{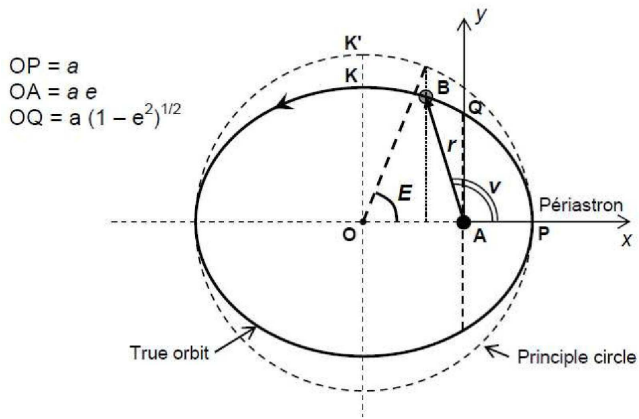}
\caption[]{
The true (B/A) relative orbit}
\label{Fig:trueorbit}
\end{figure}

The position of the B component (Figure~\ref{Fig:trueorbit}) is given by the Cartesian coordinates $x$, $y$:

\begin{equation}
\label{xycoordtrue}
x = cos E - e	= \frac{r}{a}~cos v, \quad 	
y = \sqrt{(1 - e^2)}~sin E = \frac{r}{a}~sin v,
\end{equation}

with $r$ the radius vector such as,

\begin{equation}
\label{rvcoordtrue}
r = \frac{a~(1 - e^2)}{(1 + e~cos v)}.
\end{equation}

Following the second Kepler's law, the radius vector sweeps out equal
areas in equal time differences.  The area constant $C$ is equal to
the area of the ellipse divided by the period:

\begin{equation}
\label{aeraconstant}
C = \frac{1}{2}~r^2~\frac{dv}{dt} = \frac{\pi~a^2}{P}~\sqrt{(1 - e^2)}.
\end{equation}

In the plane of the true orbit, the motion is described by Kepler's
equation:

\begin{equation}
\label{kepler}
E  - e~sin E = \frac{2~\pi~(t - T)}{P} = M,
\end{equation}

with $t$ the observation date, $E$ the eccentric anomaly, $M$ the mean
anomaly, and $v$ the true anomaly so that:

\begin{equation}
\label{trueanomaly}
tan \frac{v}{2} = \sqrt{\frac{1 + e}{1 - e}}~tan \frac{E}{2}.
\end{equation}

\subsection{The apparent orbit}
\label{subSect:apporbit}

\begin{figure}[htbp]
\centering
\includegraphics[width=0.7\textwidth]{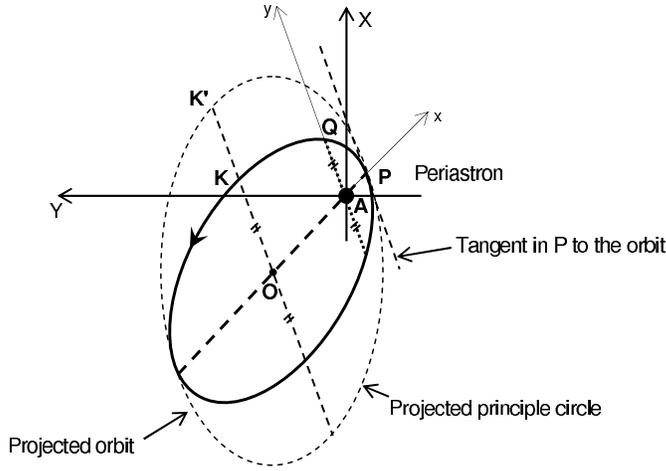}
\caption[]{
The apparent (B/A) relative orbit also called the {\it{visual orbit}}}
\label{Fig:visualorbit}
\end{figure}

The apparent orbit is the projection of the true relative (B/A) orbit
on the sky plane (see Figure~\ref{Fig:visualorbit}).  The center O of
the apparent ellipse is the center of the true orbit so that, the line
joining O and A is the projection of the semi-major axis of the true
(B/A) orbit and the orbital eccentricity is given by $e = OA/OP$.  The
line O, K' parallel to the tangent at the apparent orbit at periastron
P is the projection of the diameter of the principal circle of the
true (B/A) orbit.  The positions $X , Y$ of the B component on the sky
plane can be computed from the true orbital positions $x, y$ using:

\begin{equation}
\label{coordinates}
X = A~x + F~y,		\quad 
Y = B~x + G~y.
\end{equation}

The coefficients $A$, $B$, $F$, and $G$ are known to be the
{\it{Thiele-Innes elements}} related to the classical orbital elements
by:

\begin{equation}
\label{coeffA}
A = a~\left(\cos \omega~\cos \Omega - sin \omega~sin \Omega~\cos i \right),
\end{equation}

\begin{equation}
\label{coeffB}
B = a~\left(\cos \omega~sin \Omega + sin \omega~\cos \Omega~\cos i \right),
\end{equation}

\begin{equation}
\label{coeffF}
C = a~\left(-sin \omega~\cos \Omega - \cos \omega~sin \Omega~\cos i \right),
\end{equation}

\begin{equation}
\label{coeffG}
G = a~\left(-sin \omega~sin \Omega + \cos \omega~\cos \Omega~\cos i \right).
\end{equation}

The accumulation of measurements of the relative positions of the
components with respect to time allows to plot the apparent (B/A)
orbit. From there, several methods are used to obtain values of the
orbital elements.  Determining an orbit is a nonlinear problem whose
solution is generally obtained in two steps: computing an initial
orbit, and improvement thereof by minimizing $(O-C)$ residual using a
least-square minimization.  The geometric method of adjusting the
elliptical apparent orbit on the observed positions and determining
the orbital elements using the Thiele-Innes elements, provides a first
estimate of the orbital elements (see for
example~\citet{Heintz1978}). One can also obtain the initial orbit
using methods of automatic calculations to obtain the values of the
orbital elements that give minimum residual between the observed and
calculated positions \citep[see for example~][]{Pourbaix1994}.

\subsection{Ephemeris formulae}
\label{ephemeris}

For a given observing date $t$, the angular separation $\rho$ and the
position angle $\theta$ of a double star whose orbital elements are
known can be calculated using the ephemeris formulae.  The eccentric
anomaly $E$ is obtained by solving the Kepler's equation
(Equation~\ref{kepler}), the true anomaly $v$ and the radius vector
$r$ given by equations \ref{trueanomaly} and \ref{rvcoordtrue}
respectively.  Then, $\rho$ and $\theta$ are computed using formulae:

\begin{equation}
\label{ephem1}
tan (\theta - \Omega) = tan (v + \omega)~\cos i,
\end{equation}

\begin{equation}
\label{ephem2}
\rho = r~\frac{\cos (v + \omega)}{\cos (\theta - \Omega)}.
\end{equation}

The knowledge of some particular positions can be interesting within
the framework of the preparation of interferometric observations to
resolve a spectroscopic binary: \\

{\bf{Periastron: $v = 0^\circ$}}, 

\begin{equation}
\label{periastron}
r_P = a~(1-e), \quad 
tan (\theta_P - \Omega) = tan \omega~\cos i, \quad 
\rho_P = a~(1-e)~\frac{\cos \omega}{\cos (\theta_P-\Omega)}.
\end{equation}

{\bf{Apoastron: $v = 180^\circ$}}, 

\begin{equation}
\label{apoastron}
r_{P'} = a~(1+e), \quad 
\theta_{P'} = \theta_P + 180^\circ, \quad 
\rho_{P'} = -a~(1+e)~\frac{\cos \omega}{\cos (\theta_{P'}-\Omega)}.
\end{equation}

{\bf{Ascending node: $v = 360^\circ - \omega$}}, 

\begin{equation}
\label{node}
r_{\Omega} = a~\frac{(1-e^2)}{(1 + e~\cos \omega)}, \quad 
(\theta_{\Omega} - \Omega) = 0^\circ, \quad 
\rho_{\Omega} = a~\frac{(1-e^2)}{(1 + e~\cos \omega)}.
\end{equation}

{\bf{descending node: $v = 180^\circ - \omega$}}, 

\begin{equation}
\label{antinode}
r_{\Omega~'} = a~\frac{(1-e^2)}{(1 - e~\cos \omega)}, \quad 
(\theta_{\Omega~'} - \Omega) = 180^\circ, \quad 
\rho_{\Omega~'} = -a~\frac{(1-e^2)}{(1 - e~\cos \omega)}.
\end{equation}

It should be noted that the relative position of the component at the
time of the passage to the node gives the orientation of the line of
Nodes in the sky plane and does not depend on the orbital inclination.

\subsection{The spectroscopic orbits}
\label{spectrorbits}

If $V_{rA}$, $V_{rB}$ are the heliocentric radial velocities of
components A and B respectively and $V_{rAB}$ the radial velocity of
the center of mass AB of the binary and denoting $V_{AB/A}$ and
$V_{AB/B}$ the radial components of the orbital velocity of A and B
around the center of mass, then:

\begin{equation}
\label{Vr}
V_{rA} = V_{AB/A} + V_{rAB}, \quad 
V_{rB} = V_{AB/B} + V_{rAB},
\end{equation}

with the radial velocity $V_{rAB}$ of the center of mass or systemic
velocity $V_{sys}$:

\begin{equation}
\label{Vsys}
V_{sys} = V_{rAB} = \frac{1}{P}~\int _0 ^P V_R(t) dt,
\end{equation}

where $V_R$ is the stellar radial velocity obtained from the measurement of the spectral lines:

\begin{equation}
\label{VR}
V_R = V_{sys} + \frac{dz}{dt}.
\end{equation}

With $z$ the radius vector $r$ projected onto the line of sight:

\begin{equation}
\label{zcomp}
z = r~sin (v + \omega)~sin i.
\end{equation}

\begin{figure}[htbp]
\centering
\includegraphics[width=0.7\textwidth]{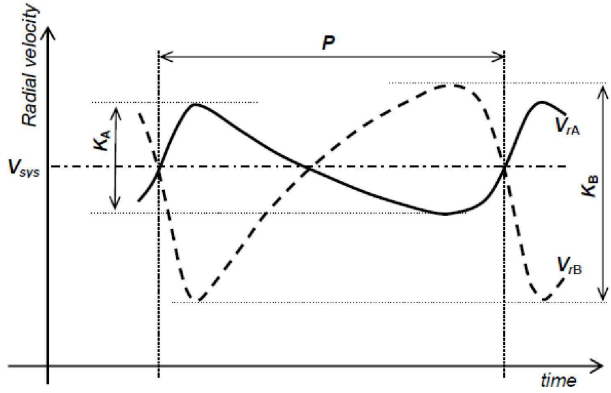}
\caption[]{
Radial velocity curves for the components of a spectroscopic binary}
\label{Fig:spectrorbits}
\end{figure}

The orbital radial velocity of each component relates to the classical
orbital elements by:

\begin{equation}
\label{spectrorbit1}
\frac{dz}{dt} = \frac{2~\pi}{P} \frac{a~sin i}{\sqrt{1-e^2}} \left[ \cos (v + \omega) + e~\cos \omega \right],
\end{equation}

where $v$ is the true anomaly, $P$ the orbital period, $a$ are the
semi-major axis $a_A$ or $a_B$, and $\omega$ the argument of the
periastron $\omega_A$ or $\omega_B$ of the (A/AB) and (B/AB) orbits
respectively.  Defining:

\begin{equation}
\label{semiamplitude}
K = \frac{2~\pi}{P} \frac{a~sin i}{\sqrt{1-e^2}}. 
\end{equation}

The maximum and minimum radial velocities are:

\begin{equation}
\label{extremaVr}
V_{max} = V_{sys} + K (1 + e~\cos \omega), \quad 	
V_{min} = V_{sys} + K (e~\cos \omega  - 1).
\end{equation}

The semi-amplitude of variation of the radial velocity being $K$ and
its median value $V_{med}$ can be estimated using:

\begin{equation}
\label{semiamplitudeK}
K = \frac{V_{max} - V_{min}}{2},	\quad 
V_{med} = \frac{V_{max} + V_{min}}{2} = V_{sys} + K~e~\cos \omega.
\end{equation}

At the date of passage to the periastron $T$:

\begin{equation}
\label{Vperiatron1}
V(T) + V(T + \frac{P}{2}) = 2~V_{med},
\end{equation}

then the radial velocities at the time of passage at periastron and
the apoastron are:

\begin{equation}
\label{Vperiatron2}
V_{per} = V_{med} + K~\cos \omega,	\quad 
V_{apo} = V_{med} - K~\cos \omega,
\end{equation}

from which it follows the determination of $\omega$ the argument of
the periastron using:

\begin{equation}
\label{omega}
K~\cos \omega = \frac{V_{per} - V_{apo}}{2}.
\end{equation}

For double line spectroscopic binaries (SB2), the measurements of
$V_{A/AB}$ and $V_{B/AB}$ are possible; for single-lined spectroscopic
binaries (SB1), only $V_{A/AB}$ measurements are possible.

For visual binaries, the relative radial velocity between the
components, $V_r~=V_{rB}~-~V_{rA}$, can be calculated from the orbital
elements of the (B/A) orbit using:

\begin{equation}
\label{visualVr}
V_r = K~\left[ \cos (v + \omega) + e~\cos \omega \right],
\end{equation}

with,

\begin{equation}
\label{Kvisualyear}
K (km/s) = 29,76~\frac{a~sin i}{P~\sqrt{1 - e^2}}, 
\end{equation}
for $P$ in year, and
\begin{equation}
\label{Kvisualday}
K (km/s) = 10870~\frac{a~sin i}{P~\sqrt{1 - e^2}}, 
\end{equation}
for $P$ in day.

For an astrometric binary orbit, the radial velocity of the
photocenter $V_{rF}$ can be computed with the same formula where the
semi-major axis $a_F$ and the argument of the periastron $\omega_F$ of
the photocenter orbit have been inserted instead. It is easy to show
that $\omega_A = \omega_F = \omega_B + 180^\circ$.

The variation of $V_r$ and $V_{rB}$ are the same and are opposite to
that of $V_{rA}$. By definition, the maxima of the $V_{rA}$ and
$V_{rB}$ curves occurs at the passage of the ascending node in the
true orbits (A/AB) and (B/AB) respectively, whereas the maximum of the
$V_r$ curve corresponds to the passage at the ascending node in the
relative orbit (B/A).

\subsubsection{Double line spectroscopic binary}
\label{subSect:SB2}

The lines of the two components are visible in the composite spectrum.
The radial velocities curves of the two components can be then
determined and allow calculating the (A/AB) and (B/AB) orbits around
the center of mass:

\begin{equation}
\label{SB2semiaxis1}
a_A~sin i = \frac{P}{2~\varpi} K_A~\sqrt{1 - e^2}, \quad 
a_B~sin i = \frac{P}{2~\varpi} K_B~\sqrt{1 - e^2}. 
\end{equation}

{\bf{Mass ratio and lower limit of the stellar masses}}

The semi-major axis of the absolute orbits being function of the
stellar masses,

\begin{equation}
\label{SB2semiaxis2}
 a_A = a~\frac{M_B}{M_A+M_B}, \quad 
 a_B = a~\frac{M_A}{M_A+M_B},
\end{equation}

with $a = a_A + a_B$ the semi-major axis of the relative (B/A) orbit.

The {\bf{mass ratio}} $\frac{M_B}{M_A}$ is equal to the ratio of the
amplitudes of the radial velocity curves:

\begin{equation}
\label{SB2massratio}
\frac{M_B}{M_A} = \frac{a_A}{a_B} = \frac{K_A}{K_B}.
\end{equation}

The determination of the spectroscopic orbit provides an estimate of
the {\bf{lower limit of the stellar masses}} of the system.

With $P$ in days, $K$ in km / s, and $M$  in solar mass it comes:

\begin{eqnarray}
\label{SB2mass1}
M_B~sin^3 i &= & 1,036~10^{-7}~K_A~\left(K_A + K_B\right)^2~P~\left(1 - e^2\right)^{3/2}, \nonumber\\
M_A~sin^3 i &= & 1,036~10^{-7}~K_B \left(K_A + K_B\right)^2~P~\left(1 - e^2\right)^{3/2}, 
\end{eqnarray}
and then,
\begin{equation}
\label{SB2mass2}
\left(M_A + M_B\right)~sin^3 i = \frac{P}{2~\pi~G}~\left(1 - e^2\right)^{3/2}~\left(K_A + K_B\right)^3.
\end{equation}

{\bf{Contribution of interferometric observations}}

If a SB2 spectroscopic binary is resolved by interferometric
observations, determination of the orbital elements of the (B/A) orbit
provides the value of the angular semi-major axis $a\arcsec$ and the
inclination $i$.  An estimation of the distance of the system can be
obtain from the ratio $a/a\arcsec = d = 1/\varpi_{orb}$, with
$\varpi_{orb}$ the {\it{orbital parallax}}.  The combination of the
mass ratio $M_B/M_A$ from the spectroscopic orbits and the total masse
$M_A + M_B$ from visual (B/A) orbit is then used to calculate the
values of the stellar masses $M_A$ and $M_B$ of each component.

\subsubsection{Single line spectroscopic binary}
\label{subsubSect:SB1}

Only the spectral lines of one of the component are recorded, this is
usually due to a large magnitude difference in between the two
components.  So, only the radial velocity curve of the brightest
component can be used to calculate the (A/AB) orbit around the center
of mass, giving:

\begin{equation}
\label{SB1semiaxis1}
a_A~sin i = \frac{P}{2~\varpi}~K_A~\sqrt{1 - e^2}.
\end{equation}

{\bf{Mass Function}}

For SB1 binary, only a combination of the stellar masses, called the
{\it{Mass Function}} can be determined from the spectroscopic orbit:

\begin{equation}
\label{SB1mass1}
f(M) = \left(M_A + M_B\right)~\left(\frac{M_B}{M_A + M_B}\right)^3~sin^3 i = \frac{P}{2~\pi~G}~\left(1 - e^2\right)^{3/2}~K_A^3.
\end{equation}

With $P$ in days, $K$ in $km/s$, and $M$  in solar masses it comes:

\begin{equation}
\label{SB1mass2}
a_A~sin i = 1375~K_A~P~\sqrt{1 - e^2} \qquad
f(M) = 1.036~10^{-7}~P~\left(1 - e^2\right)^{3/2}~K_A^3.
\end{equation}

{\bf{Contribution of interferometric observations}}

If a SB1 spectroscopic binary is resolved by interferometric
observations, determination of the (B/A) orbit provides the value of
the angular semi-major axis $a\arcsec$ and the inclination $i$.  If
the distance of the system is known (by the trigonometric parallax for
example) , the combination of parameters of spectroscopic and visual
orbits is then used to calculate the values of the stellar masses.
The linear semi-major axis (in au) $a = a\arcsec~/\varpi_{trig}$ of
the (B/A) orbit and the total mass of the system $\left(M_A +
M_B\right)$ being known, the sem-major axis of the (A/AB) orbit $a_A$
can be calculated from the $a_A~sini$ of the spectroscopic orbit
allowing estimation of the mass ratio $M_B/(MA + MB) = a_A/a$ and the
stellar masses $M_B$ and $M_A$.

It should be noted that in the case of a SB1 or SB2 spectroscopic
binary resolved as visual binary, the orbital parameter setting can be
obtained by combining astrometric and radial velocity measurements
(see for example~\citet{Pourbaix1998}).


\section[Interferometric observations]{Interferometric observation of double stars}
\label{Sect:interferobs}

The studies of the morphology of double stars still need to use
observations at high angular resolution.  Long baseline interferometry
differs from other astrometric observations of double stars by a large
gain in angular resolution. While CCD imaging achieved a resolution of
about $0.25\arcsec$ for observing visual binaries, and speckle
interferometry reaches $0.05\arcsec$ resolving as {\it{visual
binaries}} some spectroscopic binaries, the long baseline
interferometry reaches resolution of the order of $0.001\arcsec$ and
therefore it allows to resolve many spectroscopic binaries and even
some photometric binaries.

This explains why the double stars were among the earliest targets for
the interferometric observations. One can find all the interferometric
measurements of binary stars in the "{\it{Fourth Catalog of
Interferometric Measurements of Binary
Stars}}"~\footnote{http://ad.usno.navy.mil/wds/int4.html}.

\begin{figure}[htbp]
\centering
\includegraphics[width=0.3\textwidth]{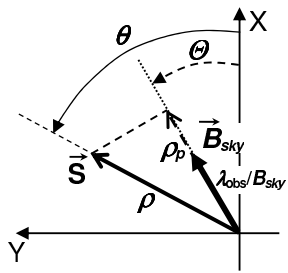}
\caption[]{
{\it{Star vector}} and {\it{baseline vector}} on the skyplane}
\label{Fig:binskyplane}
\end{figure}

For a long baseline interferometric observation of a double star, two
quantities must be taken into account (Figure~\ref{Fig:binskyplane}),
the relative position of the two components projected in the sky
plane, characterized {\it{star vector}} and the projection on the sky
of the interferometric baseline, characterized by the {\it{baseline
vector}}.

These vectors are described using the reference frame $(A, x, y, z)$
centered on the A component with two axes in the plane tangent to the
sky plane: $Ax$ points towards the North (position angle $0^\circ$),
$Ay$ points towards the East (position angle $90^\circ$).

The {\it{star vector}} is characterized by its length equal to the
angular separation of the components and its direction is given by
position angle of the two stars (see Figure~\ref{Fig:binskyplane}),
\begin{equation}
\label{vectorE}
\stackrel{\rightarrow}{S} = (S_X , S_Y) = (\rho~\cos\theta,\rho~sin\theta).
\end{equation}

The double star is also characterized by the ratio of the intensities
$I_2/I_1$ (or the difference of magnitudes $\Delta~m = m_2-m_1)$ of
these components:

\begin{equation}
\label{Intratio}
R = I_2/I_1 = 10^{-0.4\Delta m}.
\end{equation}

The {\it{baseline vector}} (see Figure~\ref{Fig:binskyplane}) is
characterized by the length $B_{sky}/\lambda$ and direction $\Theta$
of the projection on the sky of the interferometric {\it{ground
baseline}}.  The {\it{ground baseline vector}} will be assumed here
horizontal with a length and direction that are defined in a geodetic
coordinate system \footnote{The World Geodetic System (WGS) is a
standard for use in cartography, geodesy, and navigation.}.

\begin{equation}
\label{vectorBground}
\stackrel{\rightarrow}{B_{gro}} = (B_N , B_E),
\end{equation}

with $B_N$ and $B_E$ the projections toward North and East of the
{\it{ground baseline}}.

\begin{equation}
\label{vectorBsky}
\stackrel{\rightarrow}{B_{sky}} = (B_X , B_Y),
\end{equation}

with:
\begin{equation}
\label{CoordBsky}
B_X = \cos H~sin \delta~sin L - \cos \delta~\cos L, \quad
B_Y = sin H~sin L,
\end{equation}

and the polar coordinates $B_{sky} = (B_X^2 + B_Y^2)^{1/2}$ and
$tan \Theta = B_Y/B_X$.

The {\it{sky baseline vector}} depends on the latitude $L$ of the
observatory as well as the position of the target star in the sky
defined by its equatorial coordinates (right ascension $\alpha$ and
declination $\delta$) and the hour angle of observation $(H = ST
- \alpha)$.  Thus, with a fixed {\it{ground baseline}}, the {\it{sky
baseline vector}} is constantly variable depending on the time of
observation (i.e. the sidereal time $ST$).

A key parameter for interferometric observation of a double star is
the projection of the {\it{star vector}} on the direction of the
{\it{baseline vector}} (see Fig.~\ref{Fig:binskyplane}), the length of
which is:

\begin{equation}
\label{projectedrho}
\rho_p = \rho~\cos(\theta-\Theta).
\end{equation}

In the following we will see that the interferometric observation of a
double star can be done in two ways:
\begin{itemize}
\item by recording the fringe visibility of a double source;
\item by recording the fringes of each separated stellar components.
\end{itemize}

\subsection{interferometer observing a double source}
\label{subSect:doublesource}
 
\begin{figure}[htbp]
\centering
\includegraphics[width=0.9\textwidth]{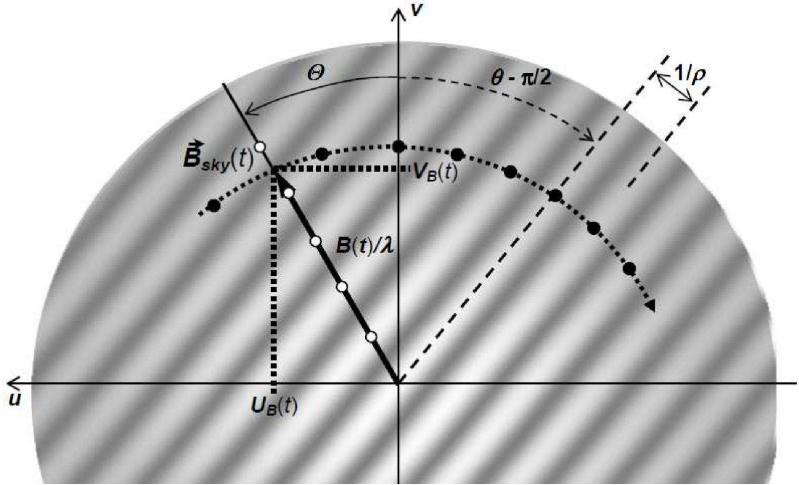}
\caption[]{
Star fringe visibility and baseline vector on the uv plane}
\label{Fig:binuvplane}
\end{figure}

For the analysis of interferometric observations, it is common to
consider vectors in Fourier space $(u, v)$, i.e. the spatial frequency
domain (see Fig.~\ref{Fig:binuvplane}).  In this space, an
interferometer characterized by a {\it{sky baseline vector}}
$\stackrel{\rightarrow}{B_{sky}}$ and observing at the wavelength
$\lambda$ must be regarded as a filter isolating the single
{\it{spatial frequency vector}} $\stackrel{\rightarrow}{f}$:

\begin{equation}
\label{vectorf}
\stackrel{\rightarrow}{f} = (f_u,f_v) = \frac{\stackrel{\rightarrow}{B_{sky}}}{\lambda},
\end{equation}

with the coordinates,

\begin{eqnarray}
\label{fcoord}
f_u &= &\frac{1}{\lambda} [-B_N~sin L~sin H - B_E~\cos H], \nonumber\\
f_v &= &\frac{1}{\lambda} [B_N~\cos L~\cos \delta + sin L~sin \delta~\cos H + B_E~sin \delta~sin H]. 
\end{eqnarray}
In this space, a double star will be represented by a {\it{corrugated sheet}} characterized by:
\begin{itemize}
\item the period of the fringes $\Lambda = 1 / \rho $;
\item the orientation of the fringes given by the position angle
      $\theta$ of the binary;
\item the amplitude of modulation of the fringes, function of the
      intensity ratio of the components and their angular diameters,
      is given by the function of visibility $V(f)$ of the source
      i.e. the Fourier transform of the light intensity distribution
      on the source:
\end{itemize}

\begin{equation}
\label{visibinmod}
V(f) = \frac{1}{1+R} \sqrt{[V_1(f)^2 + R^2~V_2(f)^2 +
2~R~V_1(f)~V_2(f)~\cos \psi]^{1/2}},
\end{equation}

with $\psi(f) = 2~\pi~\rho_p~f$, and the value of the visibility
oscillates between $V_{max}$ and $V_{min}$ values:

\begin{equation}
\label{visibinmodvar}
V_{max} = \frac{V_1(f) + R~V_2(f)}{1+R},\quad
V_{min} = \frac{V_1(f) - R~V_2(f)}{1+R}.
\end{equation}

$V(f)$ is the amplitude of the complex visibility of the source:

\begin{equation}
\label{visibin}
\tilde{V}(f) = \frac{V_1(f)}{1+R} \exp{i \frac{R~\psi}{1+R}} + \frac{V_2(f)}{1+R} \exp{-i \frac{\psi}{1+R}} = V(f)~\exp{i \Phi(f)},
\end{equation}

$V_1(f)$ and $V_2(f)$ are the visibility functions of the stellar disk
of each component. The phase $\Phi$ of the complex visibility is
calculated relatively to the position of the photometric barycenter
(photocenter) of the intensity distribution on the source, with:

\begin{equation}
\label{visibinphi}
tan \Phi(f) = \frac{V_1(f)~sin \frac{R~\Psi(f)}{1+R} - R~V_2(f)~sin \frac{\Psi(f)}{1+R}}{V_1(f) \cos \frac{R \Psi(f)}{1+R} - R~V_2(f)~\cos \frac{\Psi(f)}{1+R}}.
\end{equation}

{\bf{Observational procedures}} \\

As seen before, the earth rotation produces a rotation of the
projected {\it{baseline vector}} on the sky and then a change of the
spatial frequency measured by the interferometer as function of
time. At a given time, the length of the baseline vector is inversely
proportional to the wavelength of observation.  These findings are the
basis of the two observational procedures that can be used for the
measurement of the fringe visibility of a double star.

\begin{figure}[htbp]
\centering
\includegraphics[width=0.9\textwidth]{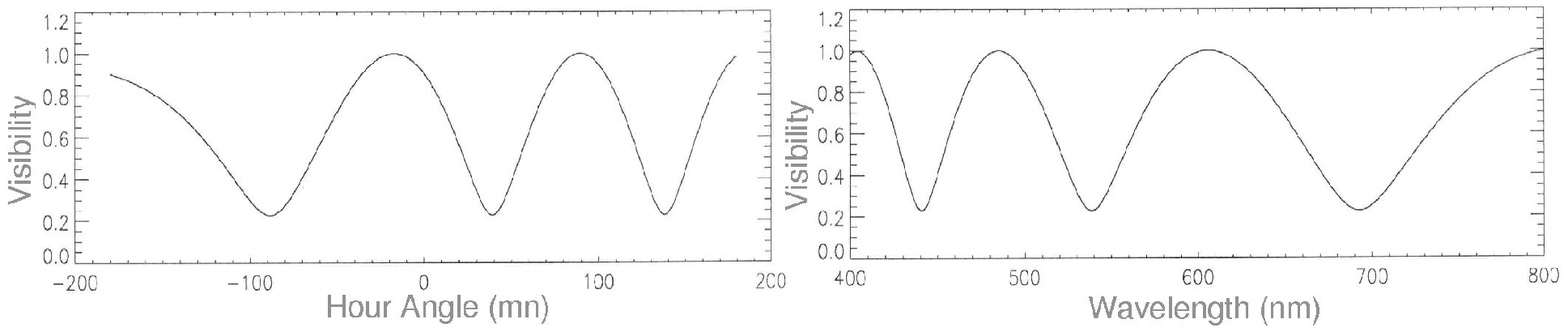}
\caption[]{
Variation of the amplitude of the visibility function for a double
star.  Up: variation of the fringe visibility as function of time
(Hour Angle) at $\lambda~=~600~nm$.  Down: variation of the fringe
visibility as function of the wavelength at $HA = 0$.

These curves was computed for a double star at $\delta = 43.75^\circ$
with $\rho = 20~mas$, $\theta = 45^\circ$, $\Delta~m = 0.5~mag$,
angular diameter of the components $\phi_{ud} = 0.1~mas$ and for a
North-South $25 m$ baseline interferometer at $L = 43.75^\circ$.  }
\label{Fig:visitimelambda}
\end{figure}

{\bf A - Variation of visibility over time.}

For given wavelength and ground baseline, the earth's rotation
produces a variation in the orientation of the baseline vector on a
resulting change with time of the angular separation projected on the
{\it{baseline vector}}.

As shown in left side of Figure~\ref{Fig:visitimelambda}, the
variation of the amplitude of the visibility as function of time
presents a series of maxima and minima.  From
equation \ref{visibinmod} it comes that:

\begin{itemize}
\item {\it{Maximum visibility}} occurs for Hour Angles $H_{max}$
      so that $\rho_p = k~\lambda/B_{sky}$
\item {\it{Minimum visibility}} occurs for Hour Angles $H_{min}$
      so that $\rho_p = (k+\frac{1}{2})~\lambda/B_{sky}$
\end{itemize}

with $k$ (integer) the interference order, the wavelength $\lambda$,
$B_{sky}$ the length of the projected baseline, and $\rho_p$ the
projected angular separation.

This method can be implemented for the interference fringes obtained
in filtered light was used by the majority of interferometers since
the late 1980s, including Mark~III (1985-1998), SUSI (since 1991) IOTA
(1993-2006), PTI (1995-2008).

\begin{figure}[htbp]
\centering
\includegraphics[width=0.9\textwidth, angle=0]{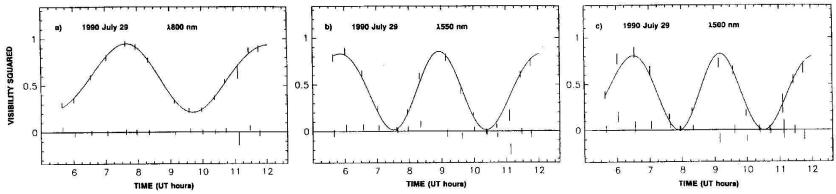}
\caption[]{
Example of observations of $\alpha$ Equulei with the MarkIII
interferometer for the night 1990 july 29.  Plot of the $V^2$ as
function of time at $\lambda~=~800, 550~nm$ for a baseline length
projected at Transit ($H = 0$) and $B_{sky} = 23,7~m$.  The error bars
on data are $\pm 1~\sigma$. The line showns the visibility computed
from the best model adjustment giving $\rho = 11,48~mas$ and $\theta =
62,4~\deg$.  (From Figure~1 in \citet{Armstrong1992}) }
\label{Fig:visimarkIII}
\end{figure}

The Figure~\ref{Fig:visimarkIII} shows early example of this type of
procedure for the observations of the double line spectroscopic binary
$\alpha$ Equulei with the MarkIII
interferometer \citep{Armstrong1992}.

\begin{figure}[htbp]
\centering
\includegraphics[width=0.95\textwidth]{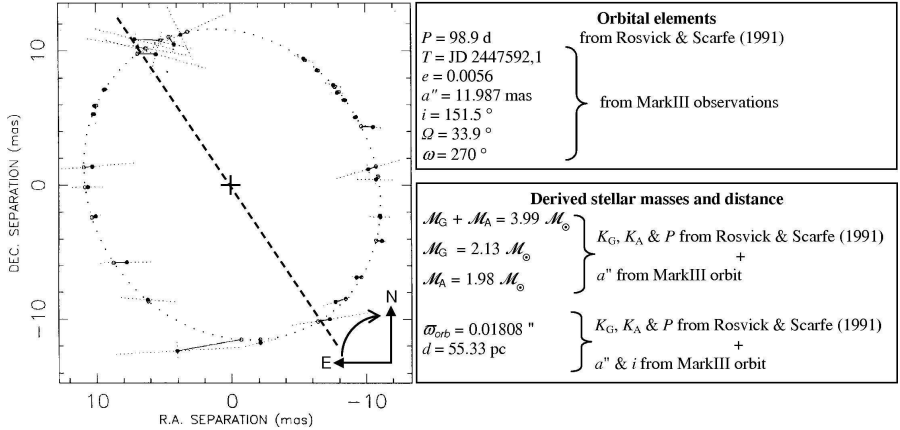}
\caption[]{
Right: Apparent orbit of $\alpha$ Equulei determined from MarkIII
interferometric observations, from Figure 2 in \citet{Armstrong1992}.
Left: Orbital elements of the relative apparent orbit, stellar masses,
and distance of $\alpha$ Equ derived from MarkIII
observations \citep{Armstrong1992}.  }
\label{Fig:markIIIorbit}
\end{figure}.

The visibility measurements at wavelengths 800, 550, 500 and 450~nm,
made during 29 nights in 1989 and 1990, were used to determine the
relative positions of the components ($\rho, \theta$) for each of
those nights and the values of the magnitude differences
$\Delta~m~(\lambda)$ in the observed spectral bands.  For the first
time, the elements of the apparent orbit of $\alpha$ Equ were obtained
by fitting the orbit on the observed positions
(Fig.\ref{Fig:markIIIorbit}).  This orbit, combined with spectroscopic
parameters of \citet{Rosvick1991} allows to determine the masses and
distance of the system.  A comparison with the stellar evolution
models was then made using the stellar masses, differences in
magnitudes, and absolute magnitudes of the components (spectral type
G5 III and A5 V) determined from interferometric observations.\\

{\bf B - Variation of the visibility as function of the wavelength.}

For fixed length of the ground baseline and a given hour angle, the
observed spatial frequency depends on the wavelength.  For a given
angular separation projected on the {\it{baseline vector}}, this
results in a variation of the value of fringe visibility as a function
of wavelength and the amplitude of the visibility presents a series of
maxima and minima over the spectrum as it is shown in right side of
Figure~\ref{Fig:visitimelambda}.  From equation \ref{visibinmod} it
comes that:

\begin{itemize}
\item {\it{Maximum visibility}} occur for the wavelengths
      $\lambda = 4.848~B_{sky}~\rho_p/k$;
\item {\it{Minimum visibility}} occur for the wavelengths
      $\lambda = 4.848~B_{sky}~\rho_p/(k+1/2)$;
\end{itemize}

with $k$ (integer) the interference order, the wavelength $\lambda$,
$B_{sky}$ the length of the projected baseline, and $\rho_p$ the
projected angular separation.

To be used, this method requires that the fringes are recorded by
dispersing the light from the star. It was implemented for the first
time on the interferometer I2T the Observatory Calern (France) in the
late 1970's and now can be used with instruments such as AMBER on the
VLTI in the IR or VEGA on CHARA in the visible.

On a basic way, these two procedures allow to determine the values of
the projected angular separation $\rho_p$ at different times of
observation. Knowing the value of the position angle $\Theta$ of the
baseline vector projected on the sky as function of the time, it is
possible to deduce the value of the angular separation $\rho$ and
position angle $\theta$ of the double star.

A more complete determination of the parameters of the double star
($\rho, \theta, R$, and also $\phi_1$ and $\phi_2$) can be obtained by
adjusting on the measured amplitude and phase of the visibility a
computed visibility curve using a model of the double star.  This
adjustment can be done using LITpro the model fitting software
developed by the "{\it{Centre Jean-Marie Mariotti}}"
(JMMC)~\footnote{http://www.mariotti.fr/litpro-page.htm}.

\begin{figure}[htbp]
\centering
\includegraphics[width=0.5\textwidth]{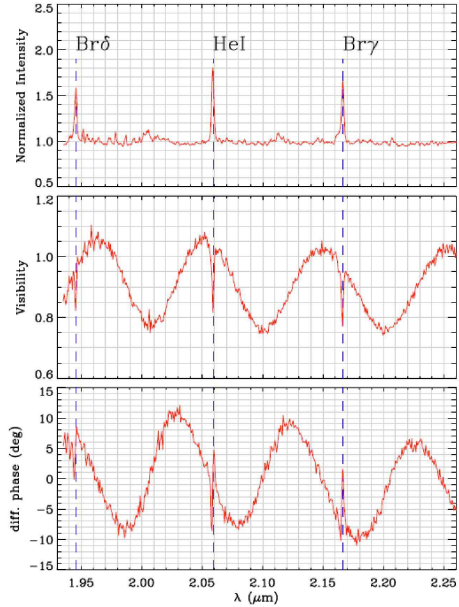}
\caption[]{
Example of a spectrally resolved fringe visiblity measurements with
VLTI/AMBER for one baseline in spectral medium resolution mode
($\frac{\lambda}{\Delta~\lambda} = 1500)$ in the K band.  Oscillations
due to the binarity of $\delta$ Sco are seen, as the three emission
lines (Br~$\delta$, HeI~$2.06~\mu m$, and Br~$\gamma$) produced by the
circumstellar disc around the primary component (known as a Be Star).
(From Figure 2 in \citet{Meilland2011}) }
\label{Fig:visidelSco}
\end{figure}

\begin{figure}[htbp]
\centering
\includegraphics[width=0.7\textwidth]{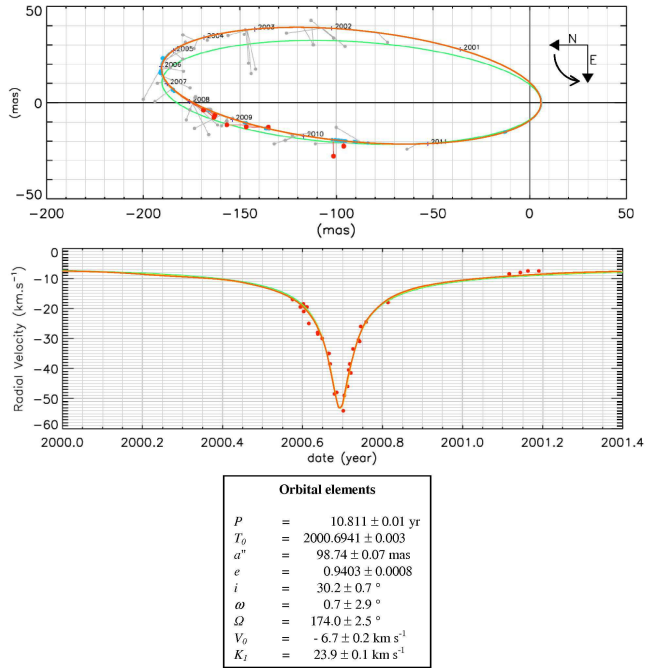}
\caption[]{
Top: Apparent orbit of $\delta$~Scorpii determined from AMBER/VLTI
interferometric observations, from Figure 3 in \citet{Meilland2011}.
Red dots are measurements derived from VLTI/AMBER data, the blue ones
from \citet{Tycner2011} and grey dots are the positions taken from de
Fourth Catalog of Interferometric Measurements of Binary Stars.  The
green and blue solid lines represent the orbits derived
by \citet{Tango2009} and \citet{Tycner2011} respectively.  The
VLTI/AMBER best-fit orbit is plotted as an orange solid line.

Middle: Radial velocitycurve of the primary component of $\delta$~Sco
around the periastron from Figure 4 in \citet{Meilland2011}.  Data
(red dots) are taken from \citet{Miroshnichenko2001}. The green and
orange lines represent \citet{Tango2009} and \citet{Meilland2011}
orbits, respectively.  Bottom: $\delta$~Sco orbital elements of the
visual and of the SB1 spectroscopic orbits as derived from VLTI/AMBER
observations \citep{Meilland2011}.  }
\label{Fig:orbitdelSco}
\end{figure}

An illustration of this observation mode is given by the observations
of the binary Be star $\delta$~Scorpii observed with the AMBER focal
instrument of the VLTI \citep{Meilland2011}.  $\delta$~Sco, resolved
as a visual binary of high eccentric orbit by speckle interferometry
since 1974, is also known as single lines spectroscopic
binary \citep{Miroshnichenko2001}.  Long baseline optical
interferometric observations were first performed in the visible with
the Sydney University Stellar Interferometer (SUSI) by measuring the
visibility as function of time \citep{Tango2009} and then with the
Navy Precision Optical Interferometer (NPOI) from polychromatic
visibility measurements \citep{Tycner2011}.  As shown in the
figure~\ref{Fig:visidelSco} the VLTI/AMBER interferometric
observations were used to measure the visibility of the fringes
dispersed in the near IR K-band around 2.2 $\mu m$.  Using the
software LITpro, the relative position of the components was obtained
by fitting the observed visibility curves by a model four free
parameters: the relative position of the components ($X, Y$), the flux
ratio ($F_1 / F_{tot}$), the diameter of the primary component
($D_1$), the secondary component assumed to be a point source.  These
new position measurements have been combined with all the published
positions and radial velocity measurements to obtain the values of
orbital parameters ($P, T_0, e, a", i, \Omega, \omega$) of the
relative orbit and the parameters ($V_0, K_1$) of the spectroscopic
orbit of the primary component (Figure~\ref{Fig:orbitdelSco}) and for
details see \citet{Meilland2011}.

.

\subsection{Double star observed as two single stars}
\label{subSect:twostars}

In this case, the double star can be described as two stars whose
positions on the sky ($\alpha_1$, $\delta_1$) and ($\alpha_2$,
$\delta_2$) corresponding to differences
$ \Delta \delta= \rho \cos \theta$ in declination and
$ \Delta \alpha= \rho sin \theta / \cos \delta_1$ in right
ascension. The differential star vector
$\stackrel{\rightarrow}{S_{12}}$ is characterized by the angular
separation of the two stars $\rho$ and the position angle $\theta$
with an angular separation projected on the baseline vector $\rho_p$.

\begin{figure}[htbp]
\includegraphics[width=0.9\textwidth]{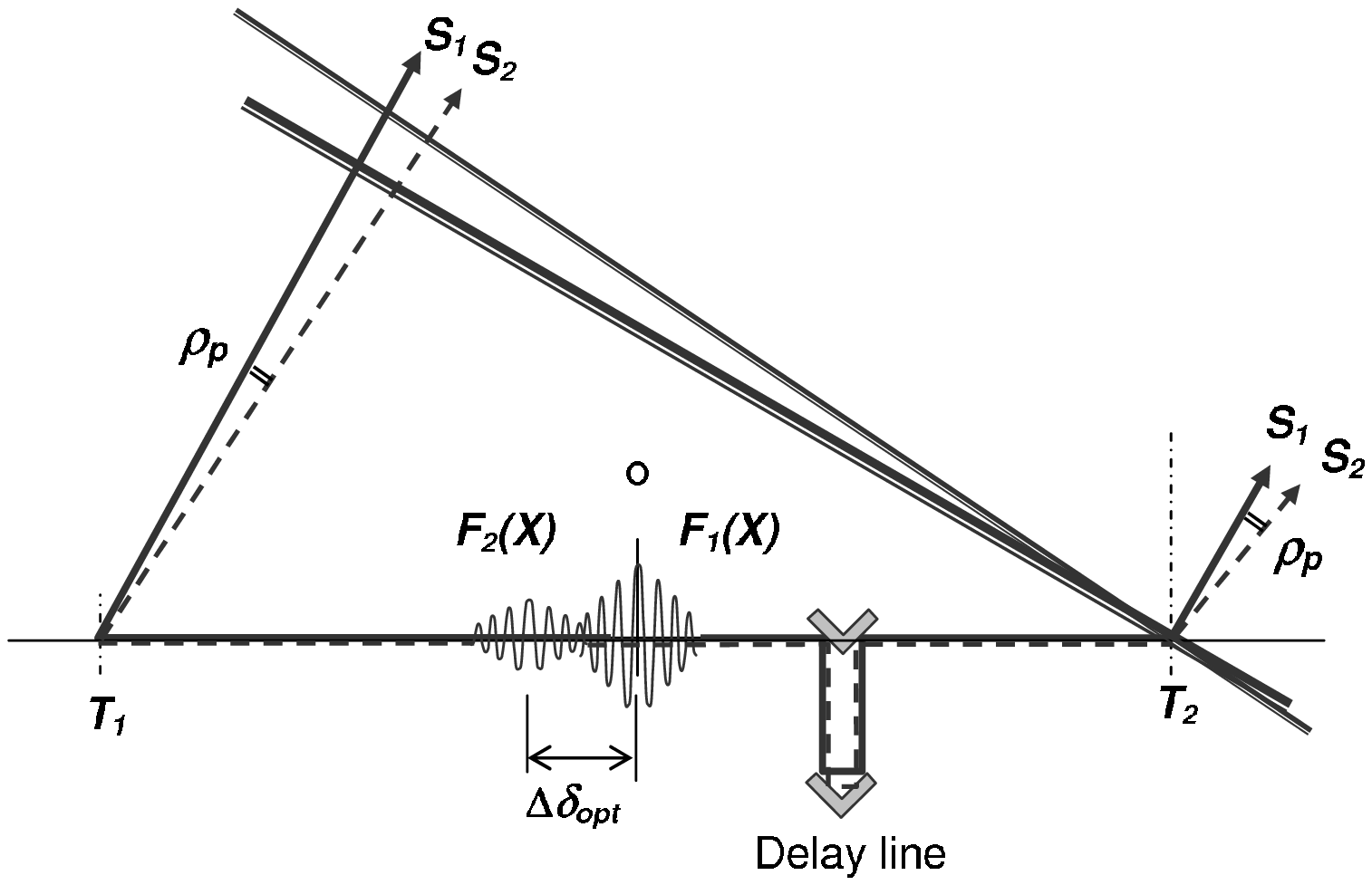}
\caption[]{Principe of the interferometric observation of a double star
 with the method of separated fringe packets.  }
\label{Fig:fringe-packets1}
\end{figure}

As shown in Fig.~\ref{Fig:fringe-packets1}, the projection of this
differential star vector $\stackrel{\rightarrow}{S_{12}}$ on the sky
baseline vector $\stackrel{\rightarrow}B_{sky}$ results in the
formation of two fringe packets $F_1$ and $F_2$ associated with each
component and separated by an optical path difference of
$\Delta \delta_{opt}$.

For a north-south horizontal baseline, we can write \citep{Kovalevsky1995}:

\begin{equation}
\label{opdbinary}
\Delta \delta_{opt}(H) = B_{gro} [ (\sin L \cos \delta_1 \sin H) \Delta \alpha - (\sin L \sin \delta_1 \cos H - \cos L \cos \delta_1) \Delta \delta],
\end{equation}

with, $\Delta \alpha = \alpha_1 - \alpha_2$ and $\Delta \delta
= \delta_1 - \delta_2$, $L$ the latitude of the interferometer and $H$
the Hour Angle of the observation.  Once the fringes are found for one
of the two stars, the search of the fringes for the second component
gives the measure of the optical paths difference between the two
fringe packets $\Delta \delta_{opt}(H)$ which are deduced the angular
separation projected on the direction of the vector base at the time
of the observation.

So far we have left out the effect of the atmospheric turbulence (the
seeing), a disturbing factor difficult to avoid during astronomical
observations.  In the case of Long Baseline Interferometry, the effect
of the atmospheric turbulence occurs primarily at two levels. i) The
random motion (tilt) and spreading (speckles) of the image making
difficult the superposition of the light beams necessary for the
formation of interference fringes.  ii) The shifts of the arrival
instants of the wavefronts on the two telescopes (piston effect)
producing random variations in the length of the optical paths in the
interferometer arms causes a random change $\sigma \delta_{opt}[atm]$
in the position $\delta_{opt}$ of the interference fringes is written
then:

\begin{equation}
\label{fringeposition}
\delta_{opt} = \stackrel{\rightarrow}{B_{sky}} . \stackrel{\rightarrow}{S} + \sigma \delta_{opt}[atm] + c,
\end{equation}

with $\stackrel{\rightarrow}B_{sky}$ the baseline vector,
$\stackrel{\rightarrow}{S}$ the star vector (see
Figure~\ref{Fig:binskyplane}), and $c$ an instrumental optical path
difference precisely measurable using a metrology device.

In principle, the simultaneous observation of the two stars eliminates
the disturbing term, but the observation procedure must satisfy the
constraints related to the properties of atmospheric turbulence,
namely:
\begin{itemize}
\item observe two stars closed enough on the sky so that the seeing can be
      considered identical on the two light beams (within the
      isoplanetic angle from about $2\arcsec- 5\arcsec$ at $\lambda =
      0.5 \mu m$ to about $10\arcsec- 30\arcsec$ at $\lambda = 2.2 \mu
      m$),
\item make the observation during a short time interval in order to consider
      the seeing as constant during the obervations (seeing lifetime
      to about $5 -10 ms$ at $\lambda = 0.5 \mu m$ to about $20-50 ms$
      at $\lambda = 2.2 \mu m$).
\end{itemize}

In these conditions, the difference in position between the two fringe
packets can be written:

\begin{equation}
\label{diffopd}
\Delta \delta_{opt} = (\delta_{opt1} - \delta_{opt2}) = \stackrel{\rightarrow}{B_{sky}}~.~(\stackrel{\rightarrow}{S_1}-\stackrel{\rightarrow}{S_2}) + (c_1 - c_2).
\end{equation}

Note that a precision of $\sigma \rho_p = \pm 10 \mu as$ in measuring
a astrometric shift $\rho_p$ requires accuracy $\sigma X = \pm 5 nm$
in the measurement of the optical path length difference
$\Delta \delta_{opt}$ between the stellar fringe packets.

The basic idea is to track the fringes simultaneously on the two stars
in order to measure their separation $\rho_p$ projected onto the
baseline vector with an accuracy of the order of a fraction of the
fringe spacing $\lambda / B_{sky}$ and to achieve narrow angle
differential astrometric observation with accuracy of the order of
$10 \mu as$ while the astrometric observations with long-focus
telescopes were around $1 mas$.

{\it Separated Fringe Packets} and {\it phase referenced
interferometry} was proposed in the 1990's \citep{Shao1992,
Quirrenbach_a1994} and allow Narrow angle interferometric measurements
of binary stars with the interferometers IOTA \citep{Dyck1995},
PTI \citep{Lane2003}, and CHARA \citep{Farrington2010}.

The {\it Separated Fringe Packets interferometry} can be used if the
stars are bright enough that the two fringe packets are detected with
an exposure time shorter than the atmospheric coherence time.  This
constraint obviously limits the sensitivity of the interferometric
observations.  The {\it phase referenced interferometry} can be used
to improve the sensitivity of the observations.  In this technique,
the fringes of a bright star are used to measure the fringe phase
fluctuations induced by atmospheric turbulence in order to correct, in
real time, the fringe phase on the fainter companion which are then
detected with an exposure time longer than the atmospheric coherence
time.

The application of this technique is illustrated by the observation of
the binary ADS 11468 AB (A 1377, HD 171779, K0 III + G9 III, $m_V$ =
5.37, $\Delta m_V$ = 0.21, $m_K$ = 2.78) made with the North-South 110
m baseline of the PTI interferometer in a spectral band of $ \lambda =
2.2~\mu m , \Delta \lambda = 0.40~\mu m $. For details
see \citet{Lane2004}.

\begin{figure}[htbp]
\centering
\includegraphics[width=0.9\textwidth]{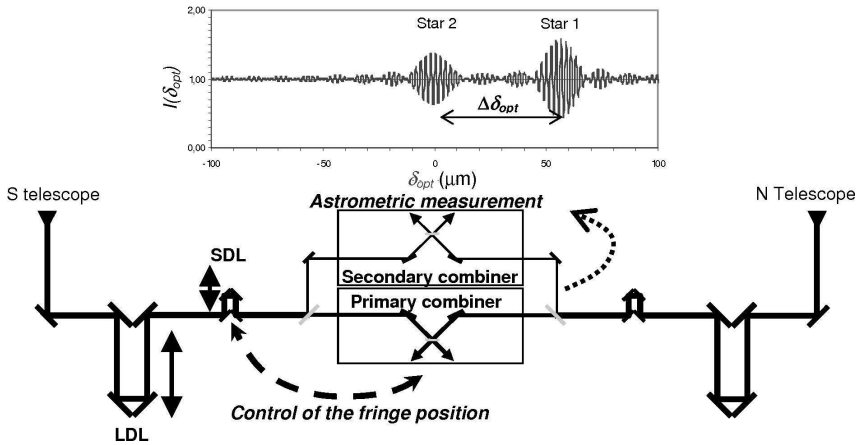}
\caption[]{Schematic diagram of the PTI instrumental configuration for Separated Fringe Packets or Phase referenced interferometric observations.
Long delay lines (LDL, optical delay $\pm 38.3 m$) and short one (SDL,
optical delay $\pm 3 cm$) placed on the southern arms of the
interferometer are active during the observation for the control of
the position of the interference fringes tracked with the primary
combiner.  The astrometric measurement is performed by the secondary
combiner.  }
\label{Fig:pti-sfp-diagram}
\end{figure}

Figure~\ref{Fig:pti-sfp-diagram} shown the schematic diagram of the
PTI instrumental configuration for Separated Fringe Packets or Phase
referenced interferometric observations \citep{Lane2003}.  The primary
fringe tracker allows to find the fringes by adjusting the optical
path difference with the long delay line (LDL) and then measures the
random fluctuations of the fringes position (or fringe phase) produced
by the atmospheric turbulence by scanning the optical path length with
an amplitude modulation of $1~\lambda$, and a frequency of $\approx
100~Hz$.  The error signal controls the short delay line (SDL) in
order to fix the position of the fringes, with a frequency of the
control loop of $\approx 10~Hz$.  The secondary fringe tracker saves
stable fringes by modulating the optical path difference with an
amplitude modulation of $\approx 500~\mu m$, a frequency $\approx
1~Hz$ and integration time 50, 100 or 250 ms. It allows to record the
two fringe packets to obtain differential astrometric measurement of
the two stars.

\begin{figure}[htbp]
\centering
\includegraphics[width=0.7\textwidth]{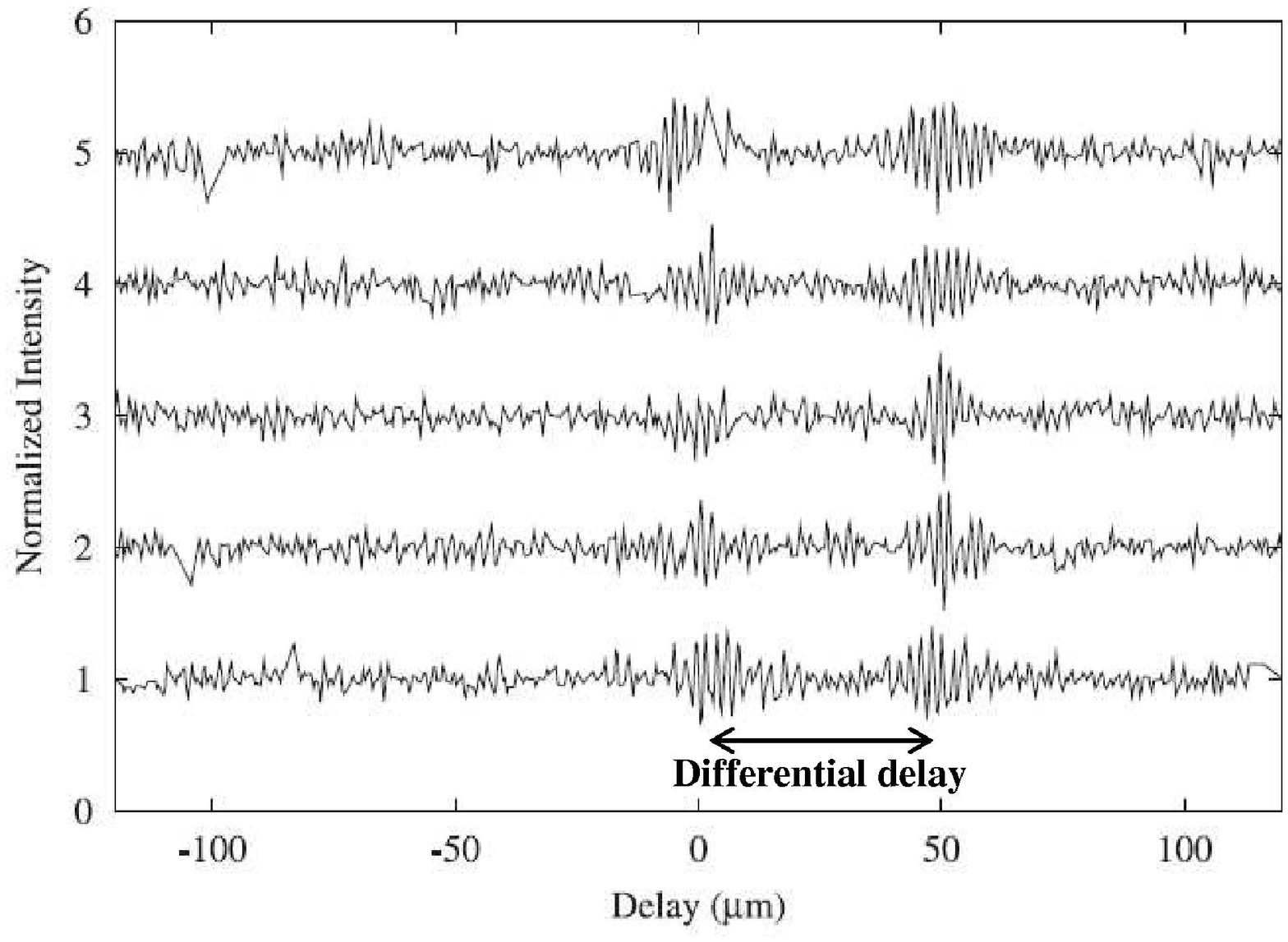}
\caption[]{Five consecutive records fringes on the visual double star
 ADS 11468 AB. The exposure time of each record is ~ 1.5 s.  (From
Figure 3 in \citet{Lane2004}) }
\label{Fig:pti-fringes-Lane2004}
\end{figure}

Figure~\ref{Fig:pti-fringes-Lane2004} shows an example of the double
fringes packets recorded by controlling the position of the fringes by
primary tracker fringe (exposure time $10~ms$, the correction optical
path difference 10 times per second) and performing the astrometric
measurement with secondary fringe tracker (amplitude $\pm 150~\mu m$,
period $3~s$). Note that the recorded signal is disturbed by the
presence of noise in part due to imperfect correction of the effects
of atmospheric turbulence.  Each scan is analyzed to obtain a
measurement of the differential delay by fitting on the data an
astrometric model with two free parameters ($\rho, \theta$) or ($X
= \Delta \alpha \cos \delta, Y = \Delta \delta$).

\begin{figure}[htbp]
\centering
\includegraphics[width=0.7\textwidth]{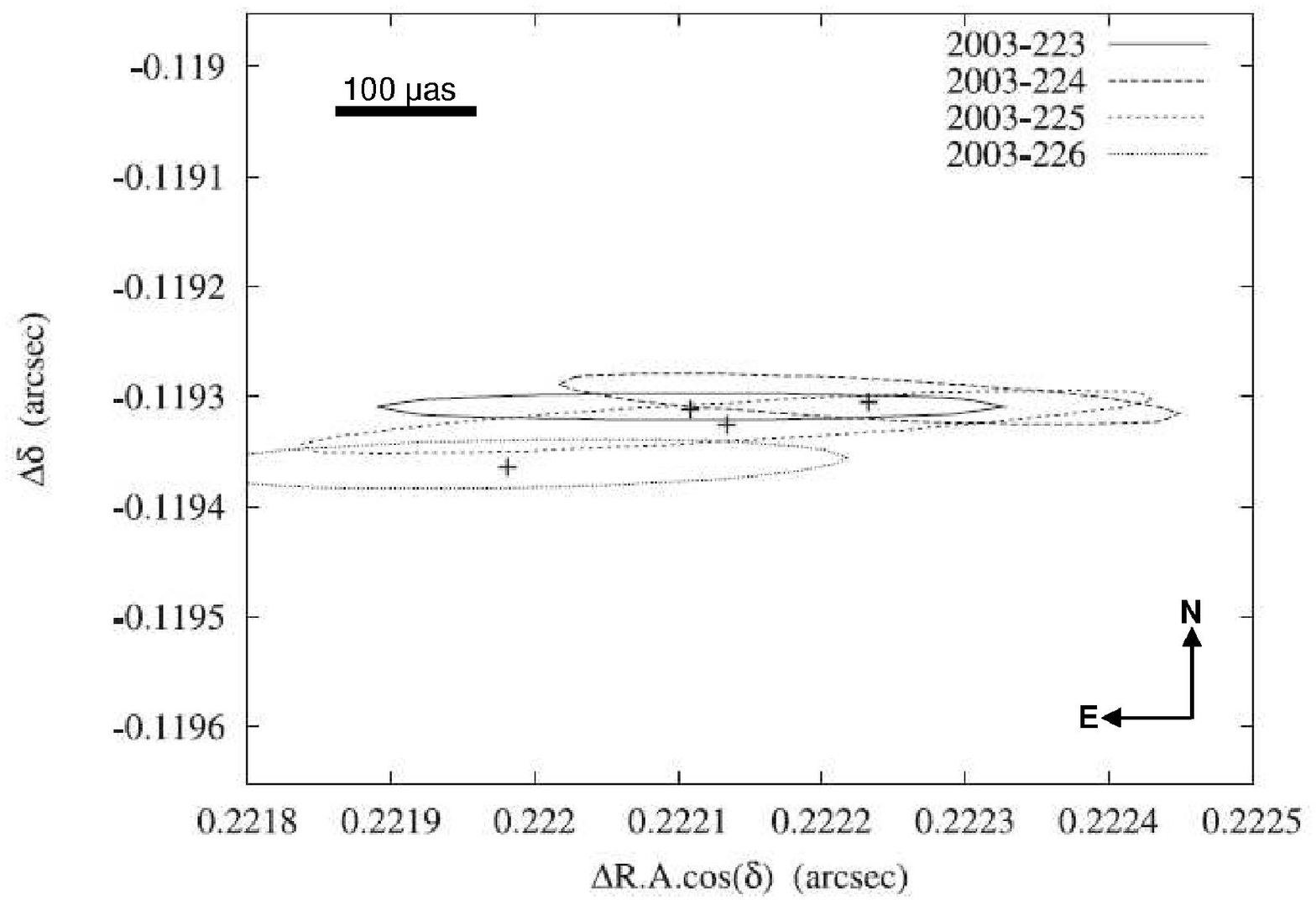}
\caption[]{Five consecutive records fringes on the visual double star
 ADS 11468 AB. The exposure time of each record is ~ 1.5 s.  (From
Figure 3 in \citet{Lane2004}) }
\label{Fig:pti-astrometry-ADS11468}
\end{figure}

Figure~\ref{Fig:pti-astrometry-ADS11468} shown that for four
consecutive nights from 11th to 14 August 2003, these observations
have identified the relative position (X, Y) between the components
(Figure Y) by fitting a simple model of interferometric binary signal
on the interferometric signal measured after correction for known
instrumental effects.  The angular separation $\rho =
0.252158~\arcsec$position and angle $\theta = 118.32~\degr$ calculated
from the mean values $<X> = -0.119324~\arcsec$ and $<Y> =
0.222138~\arcsec$ are in good agreement with the values $\rho =
0.248~\arcsec$~and $\theta = 120.9~\degr$ predicted from the orbit
period P = 191.49~years.  The uncertainties of the astrometric
differential interferometric measurements, $\sigma X \sim 144-143~\mu
as$ and $\sigma Y \sim 8-12~\mu as$ are highly anisotropic due to the
fact that the angular resolution of a North-South baseline
interferometer is actually only achieved in the direction of the
north-south component of the projected sky baseline (hence following
the declination $\delta$), while in the perpendicular direction (that
of the right ascension $\alpha$) the angular resolution remains
fundamentally limited by the short length of the east-west component
of the projected sky baseline produced by the Earth's rotation. The
main conclusion remains i.e. the errors on the interferometric
measurement are about 10 to 100 times lower than those of other
techniques.

\section[Spectra separation]{Separation of the spectra of the binary components}
\label{Sect:spectra}

Using spectro-interferometric instruments one can measure the
visibility and phase variation as function of the wavelength. As seen
in the section~\ref{subSect:doublesource}, this can be used to enhance
the (u,v) plan coverage of some observation as the spatial frequency
is defined by $\overrightarrow{B}/\lambda$. However, this assumes that
the object intensity distribution is achromatic. If the object
intensity distribution depends on the wavelength, the spectral
variations of the visibility and phase are a mix of effects from the
spatial frequency variation and the wavelength dependency of the
object.

In the simplest case of an unresolved binary star, the object
wavelength dependence is only due to the variation of the flux ratio
between the two stars. We can rewrite Equation~\ref{visibinmod}:

\begin{equation}
\label{VisiBlambda}
V(B,\lambda)=\frac{1}{1+R(\lambda)}\sqrt{1+R(\lambda)^2+2R(\lambda) \cos \left(\frac{2\pi\overrightarrow{\rho}\overrightarrow{B}}{\lambda}\right)},
\end{equation}

where $R(\lambda)$ is the wavelength dependent flux ratio between the
two components of the system. This chromatic parameter mainly
influence the variation of the visibility modulation amplitude. It can
easily be constrained using model fitting techniques. Moreover, if the
total normalized spectrum $S(\lambda)$ can be measured by the
spectro-interferometric instrument or by an other instrument, the
relative spectra of each component can be determined easily.

\begin{figure}[htbp]
\centering
\includegraphics[height=0.7\textwidth,angle=-90]{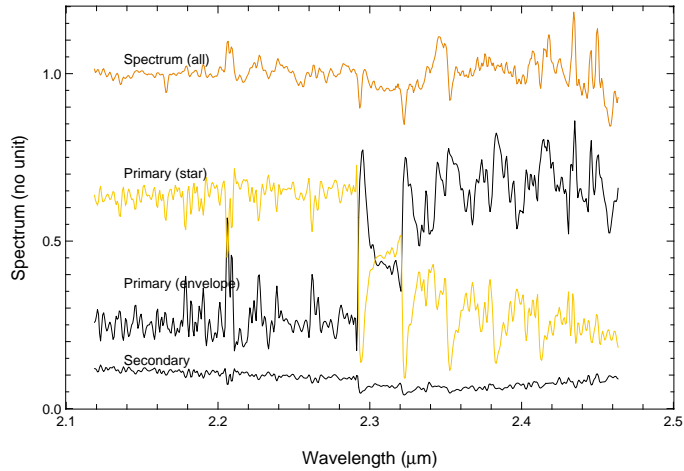}
\caption[]{Separation of the medium-resolution K band spectra HR5171\,A
 based on analysis of VLTI/AMBER data using geometric models. The
 upper orange line is the full AMBER spectrum. The yellow line is the
 spectrum from the uniform disk which its respective normalized
 flux. The contribution from the Gaussian is characterized by the
 strong CO emission lines, and the secondary flux is the bottom line
 at about 12\% level in the continuum. From \citet{Chesneau2014}.}
\label{Fig:HR5171Aspec}
\end{figure}

Using such technique \citet{Chesneau2014} managed to separate the
spectrum of the interacting binary HR\,5171\,A
(Fig~\ref{Fig:HR5171Aspec}).  In that case the authors were able to
show that the CO emission seen in the total spectrum originates from
the circumstellar environment and that the primary was showing CO
lines in absorption as expected from its inferred spectral class.

\begin{figure}[htbp]
\centering
\includegraphics[width=0.7\textwidth]{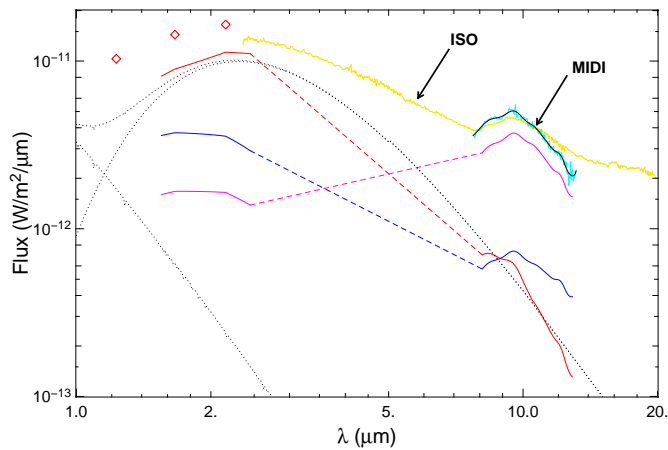}
\caption[]{The (non-de-reddened) HD\,87643 SED with the extracted fluxes from
  VLTI/AMBER and VLTI/MIDI interferometric measurements. The southern
  component flux is shown in red (top curve at 2 μm), the northern
  component flux is shown in blue (middle curve at 2 μm), and the
  resolved background flux is shown in pink (bottom curve at
  2$\mu$m). The dotted lines correspond respectively to a Kurucz
  spectrum of a B2V star and to a black-body flux at 1300\,K, for
  comparison. From \citet{Millour_b2009}}\label{Fig:HD87643}
\end{figure}

Using a similar technique one can reconstruct the spectral energy
distribution (SED) of multi-component objects using interferometric
measurements in various spectral bands and photometric or
spectro-photometric measurements. Such methods allow to put
constraints on the physical parameters of each component of the
system. A nice example showing the possibility of such technique is
the work done by \citet{Millour_b2009} on the unclassified B[e]
HD\,87643. Using AMBER and MIDI measurements the authors managed to
detect the binarity of this object surrounded by a large circumbinary
environment. They also constrained the variation of the flux ratios
between the three components of the object (the two stars and the
envelope) from 1.6 to 13$\mu$m. Finally, using some photometric
measurements from the literature for the visible and near-infrared,
and using the IRAS and MIDI spectro-photometric measurements in the
mid-infrared they manage to separate the SEDs from the primary, the
secondary and the cirumbinary envelope. As seen in
Fig~\ref{Fig:HD87643} they roughly derived the temperature of each
component, and they showed that secondary was highly embedded and that
the N-band silicate emission was mainly coming from the large
circumbinary envelope.

Finally, if the (u,v) plan coverage is well sampled and the
observations are performed with a spectro-interferometric instrument,
spectrally-resolved images can be reconstructed (see Chapter~I). In
that case the reconstructed data is very similar to those of integral
field spectroscopy but with a much higher spatial resolution. Such
technique as not been used yet on multiple stars to directly separate
the fluxes of the different component of the system, but it was
successfully applied to disentangle the stellar and circumstellar
emissions of the Be star $\phi$ Per (See Chapter~III).

\section[Interaction]{When double stars interact}
\label{Sect:interaction}

As mentioned in the introduction of this chapter, many double stars
cannot be simply regarded as two single stars orbiting around a common
center of mass. This is only the case of wide double star whose
components are subject to only their gravitational attractions but
continue their evolutions as isolated stars.  This is no longer the
case of close binaries when the separation between the two components
decreases so that the evolution of each is influenced by the presence
of the companion.

Depending on the degree of interaction between components, close
double stars are classified as detached, semi-detached, contact, and
common-envelope systems and are all considered as interacting
binaries.  The type of interaction depends on the physical properties
of stars: if at least one of the components fills its Roche lobe,
gravitational interactions mainly occurs, characterized by mass
transfer and the presence of a circumstellar or circumbinary accretion
disk. For massive binary stars, are observed not only stellar
components, but also phenomena produced by the interaction between the
hot and dense stellar winds and sometimes jet structures which are an
indication of a mass loss by the binary system.

For the study of the complex morphology of the objects, high angular
resolution observations by long baseline optical interferometry are
needed in addition to information provided by the spectroscopic and
photometric observations


\subsection{Gravitational interactions}
\label{subSect:Gravity}

\begin{figure}[htbp]
\centering
\includegraphics[width=0.6\textwidth]{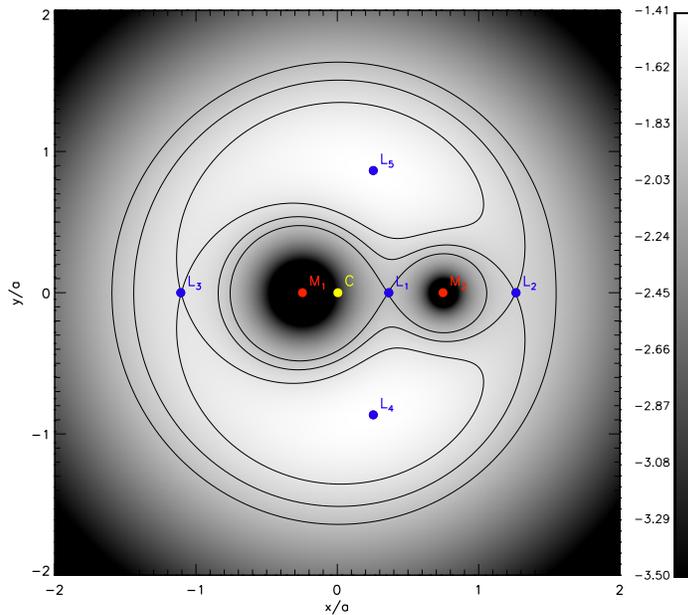}
\caption[]{Roche potential in the orbital plane for a binary system with
 a mass ratio of 3. The spatial scale is normalized by the system
 major-axis $a$. The position of the two stars are note $M_1$ and
 M$_2$, the center of Mass, C, and the five Lagrangian points L$_1$ to
 L$_5$. The equipotentials going through L$_1$ to L$_3$ are shown as a
 solid line.}
\label{Fig:RocheLobe}
\end{figure}

In close binary systems a component can have a part of its mass
accreted by the most massive component and this may affect the stellar
surfaces. To understand the possible effects, one has to take into
account the gravitational forces of both components and the
centrifugal forces induced by the rotation of the two stars around
their centre of mass. The energy potential of such system, called
Roche potential after the French Astronomer Edouard Roche, is given
by:

\begin{equation}
\label{Roche1}
\Phi(\vec{r}) = -\frac{GM_1}{\left|\vec{r}-\vec{r_1}\right|}-\frac{GM_2}{\left|\vec{r}-\vec{r_2}\right|}-\frac{1}{2}\left|\vec{\omega}\times\vec{r}\right|^2,
\end{equation}

where $M_1$ and $M_2$ are the masses of the two stars, $\vec{r_1}$ and
$\vec{r_2}$ their respective positions, and $\vec{\omega}$ the system
angular velocity. Fig~\ref{Fig:RocheLobe} illustrate the Roche
equipotentials and shows the five Lagrangian points ($L_1$ to $L_5$)
defined by $\nabla\Phi$=0. The Roche lobes of the system are defined
by the equipotential that contains $L_1$. The equipotentials of higher
energy are separated for each star, whereas those of lower energy are
merged into a single ``peanut'' shape surface.

\begin{figure}[htbp]
\centering
\includegraphics[width=0.9\textwidth]{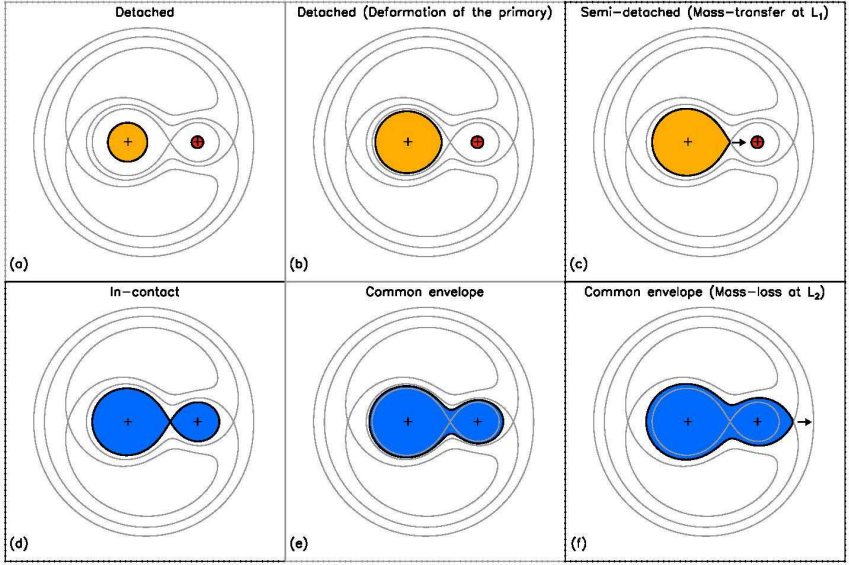}
\caption[]{Six different cases of close binary systems. The Roche
 equipotential from Fig~\ref{Fig:RocheLobe} are overplotted as solid
 grey lines. The arrows represent the possibility of mass-transfer or
 mass loss in the case of the semi-detached and common-envelope
 phases.}
\label{Fig:RocheLobe2}
\end{figure}

In the absence of other forces, these equipotentials define the
possible surfaces of the two stars. For a small-enough stars compared
to the binary separation, the gravitational effect of the companions
is negligible and the stellar surfaces are spherical as shown by the
high energy equipotential. For larger stars, i.e. lower energy, the
surface starts to be influenced by the companion and its shape starts
to differ from a sphere and to look like a pear. Nevertheless, in both
cases, the system is called ``detached'' as there is no mass-transfer
between the components. In a ``semi-detached'' system, one of the two
components completely fills its Roche lobes, and matter can easily
flow from L$_1$ into the Roche lobe of the other component, usually
spiralling down to the star, and forming an accretion disk around
it. If both stars fill their respective Roche lobe, the system will be
``in-contact'', with the contact occurring at L$_1$. Beyond that
point, the two stars can form a peanut-shaped
common-envelope. Finally, if the common-envelope reach L$_2$ , matter
can flow away from this point and form a circumbinary disk around the
system. All these possible cases are shown in
Fig~\ref{Fig:RocheLobe2}.

We note that, as components of binary systems evolve on the main
sequence and beyond, they can go through these different
phases. Moreover, the mass-transfer that occurs during the
semi-contact phase can strongly influence the stellar evolution of
both components as is the mass-loss occurring in the common-envelope
phase. Finally, the rotation of the outer layer of the stars can also
be strongly influenced by the companion. For instance, a star in a
semi-detached phase will be spin-up by tide-effect from the
companion. Moreover, outer layers of system in the in-contact or
common envelope phases are supposed to be co-rotating, i.e. rotating
at the Keplerian velocity, far above the typical velocity of evolved
stars.

\begin{figure}[htbp]
\centering
\includegraphics[width=0.9\textwidth]{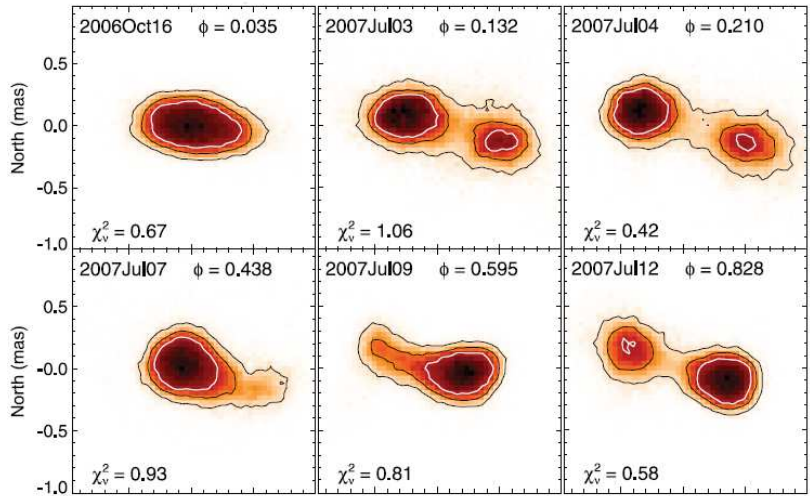}
\caption[]{Images of $\beta$ Lyr reconstructed at different epoch of
 the orbit using CHARA/MIRC data. From \citet{Zhao2008}.}
\label{Fig:betalyr}
\end{figure}

Interferometry is a well-suited technique not only to constrain the
projected orbit of interacting binaries but also to probe the geometry
of their stellar surfaces and often complex circumstellar
environment. An example of a well studied system is $\beta$ Lyr, a
semi-detached binary whose secondary is surrounded by a accretion
disk. First observations performed with the GI2T interferometer in the
H$\alpha$ emission line were published in \citet{Harmanec1996}. They
showed that the gaseous emission was stemming from an environment
larger than the binary separation, and that part of it could comes
from jet-like structures perpendicular to the orbital plane. This
hypothesis was confirmed by recent CHARA/VEGA
observations \citep{Bonneau2011}. In this paper the authors also
studied the case of another interacting binary, $\nu$ Sgr, showing
that the H$\alpha$ emission was more likely coming from a circumbinary
disk, probably fed by mass-loss flowing from L$_2$. $\beta$ Lyr was
also studied in the near-infrared using the CHARA/MIRC
instrument \citep{Zhao2008}. The author observed the system at
different epochs and managed to reconstruct images at the
corresponding phases of the orbit (see Fig~\ref{Fig:betalyr}). They
constrained the orbital parameters, and partly resolved the two
components, showing that the primary was distorted due to the Roche
lobe filling and the emission of the secondary is stemming from the
almost edge-on accretion disk.

The interferometric characterization of stellar distortion induced by
binarity is not trivial. As the distortion is small, i.e. a few
percent of the stellar diameter, very accurate measurements, and a
decent (u,v) plan coverage including baselines long enough to fully
resolve the primary, are needed to achieve this goal. For
instance, \citet{Blind2011} did not managed to derive the distortion
of the primary in SS Lep symbiotic system using VLTI/PIONIER
measurement. However, they managed to derive the mean diameter of the
primary, showing that it was not filling the Roche Lobe and that the
measured mass transfer was more likely to occur through the capture of
a part of the primary stellar wind. This result was confirmed
by \citet{Boffin2014} on three of the six symbiotic stars that they
observed with the VLTI/PIONIER instrument. For these three objects,
the authors found a filling factor of the Roche Lobe of the order of
0.5-0.6, whereas they found clues of elongation and a filling factor
close to 1 for the three others stars.

\begin{figure}[htbp]
\centering
\includegraphics[width=0.9\textwidth]{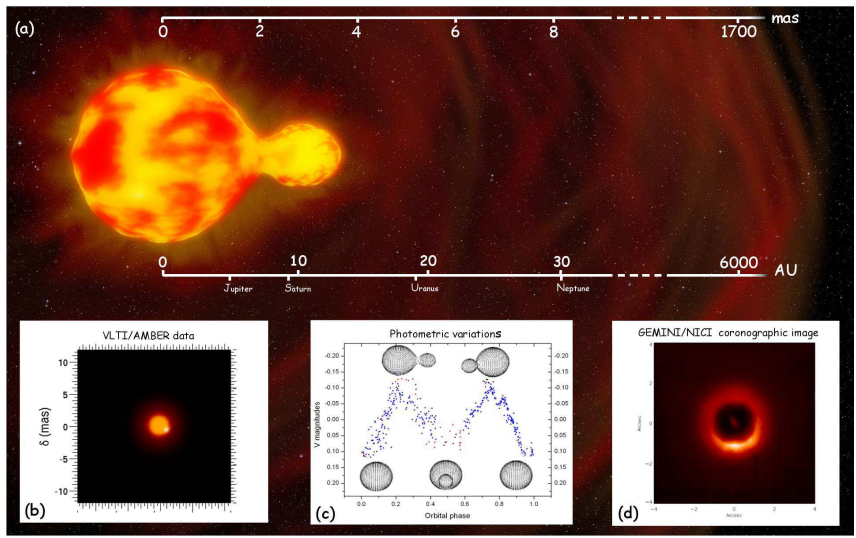}
\caption[]{(a) Artist view of HD\,5171\,A. (b) model of the VLTI/AMBER
 data. (c) photometric variations. (d) GEMINI/NICI large scale
 coronographic image. From \citep{Chesneau2014}.}
\label{Fig:HR5171A}
\end{figure}

Interferometry also allowed to discover the binary nature of one of
the most extended star of the galaxy, the yellow hypergiant
HR\,5171\,A \citep{Chesneau2014}. VLTI/AMBER data revealed the
presence of a very close companion that was located at the limb of the
primary disk during the observations (see Fig~\ref{Fig:HR5171A}). A
detailed study of the photometric variation of the stars confirmed the
binarity of the system that the star was in a in-contact for common
envelope phase. Finally, a careful analysis of the spectrum, confirmed
that the primary surface was rotating close to the breakup velocity as
expected for close binaries in common-envelope phase.

Tide effects from a close companion can also affect the circumstellar
environment of the primary. For instance, this is the case of some
classical Be stars. If the ellipticity is small and the stellar
equator and orbit are co-planar, the companion might simply truncate
the disk as for $\zeta$ Tau \citep{Quirrenbach_b1994} or $\phi$
Per \citep{Mourard2014}. For a highly elliptic orbit with a
close-enough periastron, the circumstellar disk around the primary is
likely to be perturbed by the companion during the periastron
passage. For instance, this is the case of $\delta$
Sco \citep{Che2012, Meilland2013}. Finally, if the orbital and stellar
equatorial plane are not coplanar, the tide effects can torque the
disk.

Disk and companion interaction are not fully understood yet. For
instance, although most of the stars showing the B[e] phenomenon are
found to be binaries, the mechanisms leading to the formation of a
dense equatorial disk is not yet understood. In the case of these hot
and massive stars, one as to take into account, not only the
gravitational interaction between the components, but also some
possible wind-wind collision or wind capture by the companion.

HD\,62623, the coldest star showing the B[e] phenomenon was observed
with the VLTI instruments MIDI \citep{Meilland2010} and
AMBER \citep{Millour2011}. These observations show that the star is
surrounded by a stratified gaseous and dusty disk. Constraints on the
inclination angle show that the star is rotating too slow for the
bi-stable mechanism \citep{Lamers1991} to explain the disk formation,
but too fast for a supergiant star. The companion, not detected in the
interferometric data but located in the disk, could have spin-up the
primary as for HR\,5171\,A. The disk may be a residual from a previous
interaction.

\subsection{Shock interactions}
\label{subSect:shock}

\paragraph{Wind interactions}

When two hot massive stars in a pair interact, there is no accretion,
due to the fast and dense radiation-driven winds. Indeed, these winds
are accelerated by the ``line blanketing'' effect (i.e. the numerous
interactions of photons with the many broadened lines the gaz offers
to the luminous flux from the star). Instead, their winds crash onto
each other, producing a bow-shock and a wealth of associated
phenomena.

Depending on the energy of the shock, UV, X-Rays or even sometimes
Gamma-Rays can be emitted in the region of the shock. These energetic
radiations participate to the ionization of the surrounding gazeous
regions, hence producing free-free emission (continuum) and line
emission. This additional emission may be detected by means of
interferometric observations, e.g. for $\gamma$ Velorum for the
free-free emission \citep{Millour2007}. In other cases, there
is only indirect evidence of the shock due to the highly eccentric
orbit of the secondary star, or the presence of dust in the system
which veils some of the shock clues. This is for example the case of
Eta Carin\ae.

\begin{figure}[htbp]
\begin{center}
\begin{minipage}[t]{.25\linewidth}
\vspace{0pt}
\centering
\includegraphics[width=4cm,origin=t,angle=0]{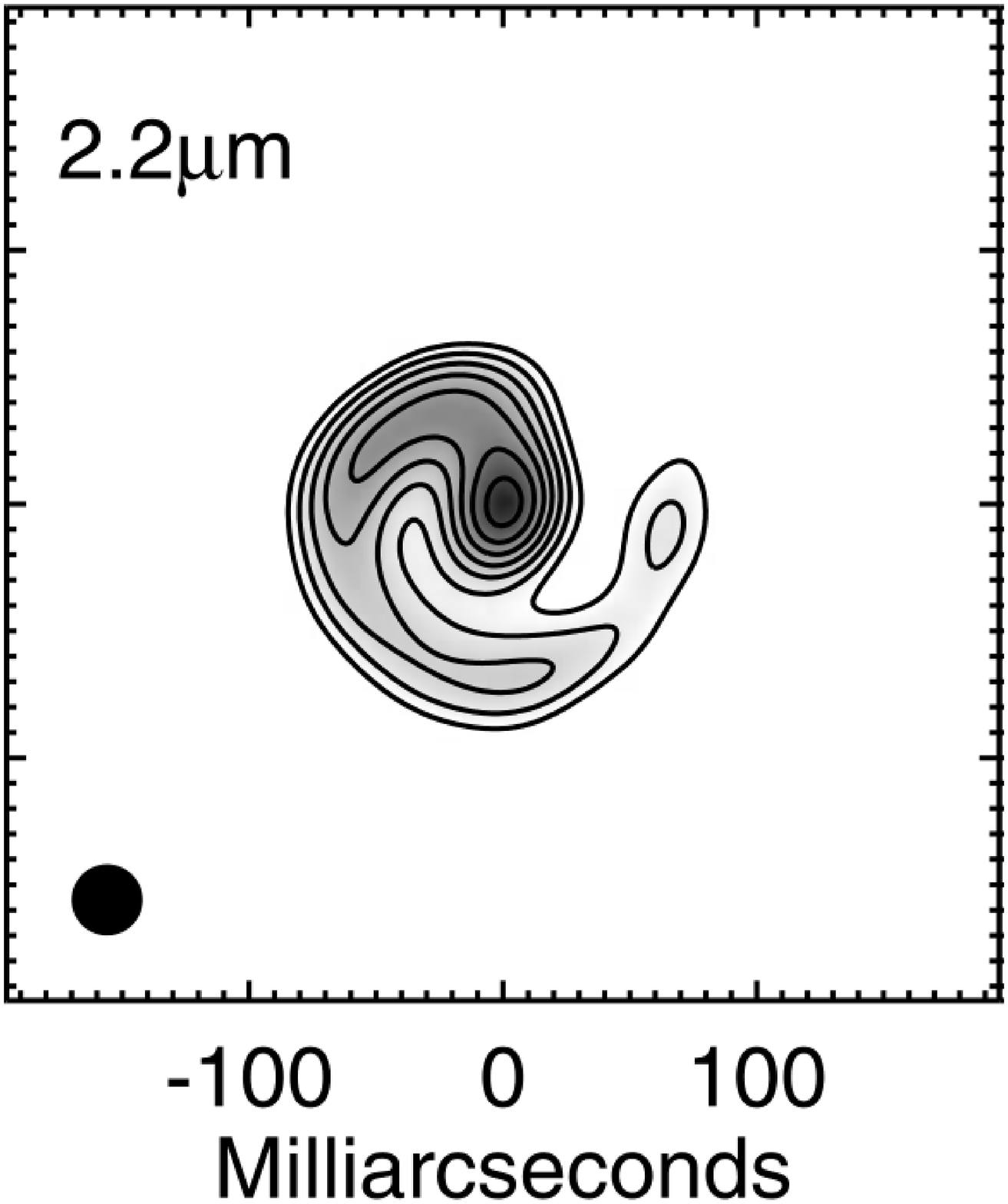}
\end{minipage}%
\hspace{1.5cm}
\begin{minipage}[t]{.25\linewidth}
\vspace{0pt}
\centering
\includegraphics[width=5.2cm,origin=t,angle=0]{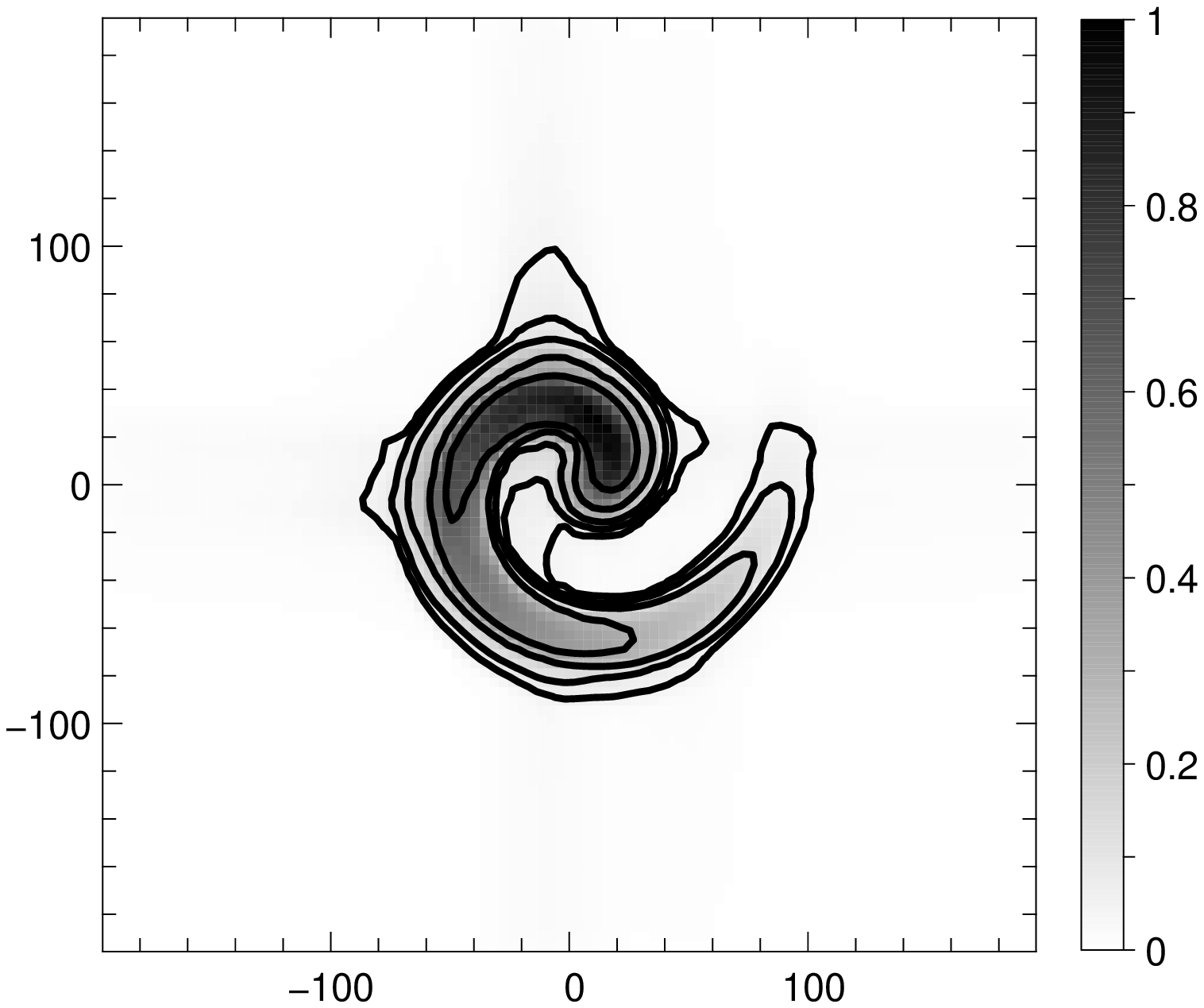}
\end{minipage}
\end{center}
\caption[]{ {\bf Left:} Figure from \citep{Monnier2007}
  showing the ``pinwheel'' nebula around the Wolf-Rayet star WR104 in
  the near infrared. {\bf Right:} The same model as in
  \citet{Millour_a2009} with parameters that make the image similar to WR104
  image.}
\label{Fig:wr104andmodel}
\end{figure}

Finally, the region of the shock is a heavily turbulent, dense, and slow
region between the winds of the two stars, which are somewhat less
dense medium, faster and a strong radiation source. Somewhere in the
shock, due to these ingredients, the metallic ions condense into dust,
which form an expanding plume of material. This plume has a very
characteristic shape (see Fig.~\ref{Fig:wr104andmodel}) named a
``pinwheel'' nebula. This is for example the case of the candidate
Gamma Ray burst binary star WR104 \citep{Monnier2007}, or
another dusty Wolf-Rayet star WR140 \citep{Monnier2002}. For
such system, the naked binary star model does not match the data, and
specific models like in \citet{Harries2004} or
\citep{Millour_a2009} need to be developed.

\paragraph{Cataclysmic binaries}

Cataclysmic binaries are made of a white dwarf (WD, receiver) and a
main-sequence or giant star (donnor). A mass transfer (accretion) from
the donnor to the receiver occurs through the Roche lobe contact at
the Lagrange points.

As a consequence, the WD surface gains mass until it reaches the
thermonuclear ignition mass, causing it to explode explode in a
thermonuclear runaway (TNR). This results in a brightening of 7 to 15
magnitudes called a Nova. Once the explosion has vanished, the cycle
of mass-gaining/TNR can start again with a typical period of
$10^4$\,years.

At each growth/TNR cycle, the white dwarf gains a little mass, making
accelerate the cycle time. Therefore, some novae have cycles with
periods as small as 10 years. These novae are called recurrent
novae. The very short timescale of recurrence indicates a WD close to
the Chandrasekhar limit, making it a candidate for Type Ia supernovae.

The work done on Novae with interferometry is well-synthesised in the
review of \citet{Chesneau2012}. It started in 1992 with Nova Cygni
1992 observed with the Mk III interferometer, and until now, almost
all bright novae were observed using one or the other interferometric
facilities available (VLTI, PTI, CHARA, etc.).

The main conclusion from these studies is that the novae ejecta are
not spherical, as was previously considered. This is even the case
just a few days after the outburst \citep{Chesneau2007, Chesneau2011},
indicating that the ejecta shaping occurs already at the earliest
phases of the nova explosion and not with later interactions with the
ISM. This has strong implications on the determination of the distance
of the system using spectroscopic parallax estimates, as demonstrated
in \citet{Schaefer2014}.

%

\begin{thebibliography}{47}
\expandafter\ifx\csname natexlab\endcsname\relax\def\natexlab#1{#1}\fi

\bibitem[{{Armstrong} {et~al.}(1992){Armstrong}, {Mozurkewich}, {Vivekanand},
  {Simon}, {Denison}, {Johnston}, {Pan}, {Shao}, \& {Colavita}}]{Armstrong1992}
{Armstrong}, J.~T., {Mozurkewich}, D., {Vivekanand}, M., {et~al.} 1992, \aj,
  104, 241

\bibitem[{{Blind} {et~al.}(2011){Blind}, {Boffin}, {Berger}, {Le Bouquin},
  {M{\'e}rand}, {Lazareff}, \& {Zins}}]{Blind2011}
{Blind}, N., {Boffin}, H.~M.~J., {Berger}, J.-P., {et~al.} 2011, \aap, 536, A55

\bibitem[{{Boffin} {et~al.}(2014){Boffin}, {Hillen}, {Berger}, {Jorissen},
  {Blind}, {Le Bouquin}, {Miko{\l}ajewska}, \& {Lazareff}}]{Boffin2014}
{Boffin}, H.~M.~J., {Hillen}, M., {Berger}, J.~P., {et~al.} 2014, \aap, 564, A1

\bibitem[{{Bonneau} {et~al.}(2011){Bonneau}, {Chesneau}, {Mourard},
  {B{\'e}rio}, {Clausse}, {Delaa}, {Marcotto}, {Perraut}, {Roussel}, {Spang},
  {Stee}, {Tallon-Bosc}, {McAlister}, {ten Brummelaar}, {Sturmann}, {Sturmann},
  {Turner}, {Farrington}, \& {Goldfinger}}]{Bonneau2011}
{Bonneau}, D., {Chesneau}, O., {Mourard}, D., {et~al.} 2011, \aap, 532, A148

\bibitem[{{Che} {et~al.}(2012){Che}, {Monnier}, {Tycner}, {Kraus}, {Zavala},
  {Baron}, {Pedretti}, {ten Brummelaar}, {McAlister}, {Ridgway}, {Sturmann},
  {Sturmann}, \& {Turner}}]{Che2012}
{Che}, X., {Monnier}, J.~D., {Tycner}, C., {et~al.} 2012, \apj, 757, 29

\bibitem[{{Chesneau} \& {Banerjee}(2012)}]{Chesneau2012}
{Chesneau}, O. \& {Banerjee}, D.~P.~K. 2012, Bulletin of the Astronomical
  Society of India, 40, 267

\bibitem[{{Chesneau} {et~al.}(2011){Chesneau}, {Meilland}, {Banerjee}, {Le
  Bouquin}, {McAlister}, {Millour}, {Ridgway}, {Spang}, {ten Brummelaar},
  {Wittkowski}, {Ashok}, {Benisty}, {Berger}, {Boyajian}, {Farrington},
  {Goldfinger}, {Merand}, {Nardetto}, {Petrov}, {Rivinius}, {Schaefer},
  {Touhami}, \& {Zins}}]{Chesneau2011}
{Chesneau}, O., {Meilland}, A., {Banerjee}, D.~P.~K., {et~al.} 2011, \aap, 534,
  L11

\bibitem[{{Chesneau} {et~al.}(2014){Chesneau}, {Meilland}, {Chapellier},
  {Millour}, {van Genderen}, {Naz{\'e}}, {Smith}, {Spang}, {Smoker}, {Dessart},
  {Kanaan}, {Bendjoya}, {Feast}, {Groh}, {Lobel}, {Nardetto}, {Otero},
  {Oudmaijer}, {Tekola}, {Whitelock}, {Arcos}, {Cur{\'e}}, \&
  {Vanzi}}]{Chesneau2014}
{Chesneau}, O., {Meilland}, A., {Chapellier}, E., {et~al.} 2014, \aap, 563, A71

\bibitem[{{Chesneau} {et~al.}(2007){Chesneau}, {Nardetto}, {Millour}, {Hummel},
  {Domiciano de Souza}, {Bonneau}, {Vannier}, {Rantakyr{\"o}}, {Spang},
  {Malbet}, {Mourard}, {Bode}, {O'Brien}, {Skinner}, {Petrov}, {Stee},
  {Tatulli}, \& {Vakili}}]{Chesneau2007}
{Chesneau}, O., {Nardetto}, N., {Millour}, F., {et~al.} 2007, \aap, 464, 119

\bibitem[{{Davis}(2007)}]{Davis2007}
{Davis}, J. 2007, in IAU Symposium, Vol. 240, IAU Symposium, ed. W.~I.
  {Hartkopf}, P.~{Harmanec}, \& E.~F. {Guinan}, 45--53

\bibitem[{{Duch{\^e}ne} \& {Kraus}(2013)}]{Duchene2013}
{Duch{\^e}ne}, G. \& {Kraus}, A. 2013, \araa, 51, 269

\bibitem[{{Dyck} {et~al.}(1995){Dyck}, {Benson}, \& {Schloerb}}]{Dyck1995}
{Dyck}, H.~M., {Benson}, J.~A., \& {Schloerb}, F.~P. 1995, \aj, 110, 1433

\bibitem[{{Farrington} {et~al.}(2010){Farrington}, {ten Brummelaar}, {Mason},
  {Hartkopf}, {McAlister}, {Raghavan}, {Turner}, {Sturmann}, {Sturmann}, \&
  {Ridgway}}]{Farrington2010}
{Farrington}, C.~D., {ten Brummelaar}, T.~A., {Mason}, B.~D., {et~al.} 2010,
  \aj, 139, 2308

\bibitem[{{Finsen}(1934)}]{Finsen1934}
{Finsen}, W.~S. 1934, Circular of the Union Observatory Johannesburg, 91, 23

\bibitem[{{Guinan} {et~al.}(2007){Guinan}, {Harmanec}, \&
  {Hartkopf}}]{guinan2007}
{Guinan}, E.~F., {Harmanec}, P., \& {Hartkopf}, W. 2007, in IAU Symposium, Vol.
  240, IAU Symposium, ed. W.~I. {Hartkopf}, P.~{Harmanec}, \& E.~F. {Guinan},
  5--16

\bibitem[{{Harmanec} {et~al.}(1996){Harmanec}, {Morand}, {Bonneau}, {Jiang},
  {Yang}, {Guinan}, {Hall}, {Mourard}, {Hadrava}, {Bozic}, {Sterken},
  {Tallon-Bosc}, {Walker}, {McCook}, {Vakili}, {Stee}, \& {Le
  Contel}}]{Harmanec1996}
{Harmanec}, P., {Morand}, F., {Bonneau}, D., {et~al.} 1996, \aap, 312, 879

\bibitem[{{Harries} {et~al.}(2004){Harries}, {Monnier}, {Symington}, \&
  {Kurosawa}}]{Harries2004}
{Harries}, T.~J., {Monnier}, J.~D., {Symington}, N.~H., \& {Kurosawa}, R. 2004,
  \mnras, 350, 565

\bibitem[{{Hartkopf} {et~al.}(2001){Hartkopf}, {McAlister}, \&
  {Mason}}]{Hartkopf2001}
{Hartkopf}, W.~I., {McAlister}, H.~A., \& {Mason}, B.~D. 2001, \aj, 122, 3480

\bibitem[{{Heintz}(1978)}]{Heintz1978}
{Heintz}, W.~D. 1978, Geophysics and Astrophysics Monographs, 15

\bibitem[{{Kovalevsky}(1995)}]{Kovalevsky1995}
{Kovalevsky}, J. 1995, {Modern Astrometry} (Springer-Verlag)

\bibitem[{{Lamers} \& {Pauldrach}(1991)}]{Lamers1991}
{Lamers}, H.~J.~G. \& {Pauldrach}, A.~W.~A. 1991, \aap, 244, L5

\bibitem[{{Lane} \& {Colavita}(2003)}]{Lane2003}
{Lane}, B.~F. \& {Colavita}, M.~M. 2003, \aj, 125, 1623

\bibitem[{{Lane} \& {Muterspaugh}(2004)}]{Lane2004}
{Lane}, B.~F. \& {Muterspaugh}, M.~W. 2004, \apj, 601, 1129

\bibitem[{{McAlister}(2007)}]{McAlister2007}
{McAlister}, H.~A. 2007, in IAU Symposium, Vol. 240, IAU Symposium, ed. W.~I.
  {Hartkopf}, P.~{Harmanec}, \& E.~F. {Guinan}, 35--44

\bibitem[{{Meilland} {et~al.}(2011){Meilland}, {Delaa}, {Stee}, {Kanaan},
  {Millour}, {Mourard}, {Bonneau}, {Petrov}, {Nardetto}, {Marcotto}, {Roussel},
  {Clausse}, {Perraut}, {McAlister}, {ten Brummelaar}, {Sturmann}, {Sturmann},
  {Turner}, {Ridgway}, {Farrington}, \& {Goldfinger}}]{Meilland2011}
{Meilland}, A., {Delaa}, O., {Stee}, P., {et~al.} 2011, \aap, 532, A80

\bibitem[{{Meilland} {et~al.}(2010){Meilland}, {Kanaan}, {Borges Fernandes},
  {Chesneau}, {Millour}, {Stee}, \& {Lopez}}]{Meilland2010}
{Meilland}, A., {Kanaan}, S., {Borges Fernandes}, M., {et~al.} 2010, \aap, 512,
  A73

\bibitem[{{Meilland} {et~al.}(2013){Meilland}, {Stee}, {Spang}, {Malbet},
  {Massi}, \& {Schertl}}]{Meilland2013}
{Meilland}, A., {Stee}, P., {Spang}, A., {et~al.} 2013, \aap, 550, L5

\bibitem[{{Millour} {et~al.}(2009{\natexlab{a}}){Millour}, {Chesneau}, {Borges
  Fernandes}, {Meilland}, {Mars}, {Benoist}, {Thi{\'e}baut}, {Stee}, {Hofmann},
  {Baron}, {Young}, {Bendjoya}, {Carciofi}, {Domiciano de Souza}, {Driebe},
  {Jankov}, {Kervella}, {Petrov}, {Robbe-Dubois}, {Vakili}, {Waters}, \&
  {Weigelt}}]{Millour_b2009}
{Millour}, F., {Chesneau}, O., {Borges Fernandes}, M., {et~al.}
  2009{\natexlab{a}}, \aap, 507, 317

\bibitem[{{Millour} {et~al.}(2009{\natexlab{b}}){Millour}, {Driebe},
  {Chesneau}, {Groh}, {Hofmann}, {Murakawa}, {Ohnaka}, {Schertl}, \&
  {Weigelt}}]{Millour_a2009}
{Millour}, F., {Driebe}, T., {Chesneau}, O., {et~al.} 2009{\natexlab{b}}, \aap,
  506, L49

\bibitem[{{Millour} {et~al.}(2011){Millour}, {Meilland}, {Chesneau}, {Stee},
  {Kanaan}, {Petrov}, {Mourard}, \& {Kraus}}]{Millour2011}
{Millour}, F., {Meilland}, A., {Chesneau}, O., {et~al.} 2011, \aap, 526, A107

\bibitem[{{Millour} {et~al.}(2007){Millour}, {Petrov}, {Chesneau}, {Bonneau},
  {Dessart}, {Bechet}, {Tallon-Bosc}, {Tallon}, {Thi{\'e}baut}, {Vakili},
  {Malbet}, {Mourard}, {Antonelli}, {Beckmann}, {Bresson}, {Chelli},
  {Dugu{\'e}}, {Duvert}, {Gennari}, {Gl{\"u}ck}, {Kern}, {Lagarde}, {Le
  Coarer}, {Lisi}, {Perraut}, {Puget}, {Rantakyr{\"o}}, {Robbe-Dubois},
  {Roussel}, {Tatulli}, {Weigelt}, {Zins}, {Accardo}, {Acke}, {Agabi},
  {Altariba}, {Arezki}, {Aristidi}, {Baffa}, {Behrend}, {Bl{\"o}cker},
  {Bonhomme}, {Busoni}, {Cassaing}, {Clausse}, {Colin}, {Connot},
  {Delboulb{\'e}}, {Domiciano de Souza}, {Driebe}, {Feautrier}, {Ferruzzi},
  {Forveille}, {Fossat}, {Foy}, {Fraix-Burnet}, {Gallardo}, {Giani}, {Gil},
  {Glentzlin}, {Heiden}, {Heininger}, {Hernandez Utrera}, {Hofmann}, {Kamm},
  {Kiekebusch}, {Kraus}, {Le Contel}, {Le Contel}, {Lesourd}, {Lopez}, {Lopez},
  {Magnard}, {Marconi}, {Mars}, {Martinot-Lagarde}, {Mathias}, {M{\`e}ge},
  {Monin}, {Mouillet}, {Nussbaum}, {Ohnaka}, {Pacheco}, {Perrier}, {Rabbia},
  {Rebattu}, {Reynaud}, {Richichi}, {Robini}, {Sacchettini}, {Schertl},
  {Sch{\"o}ller}, {Solscheid}, {Spang}, {Stee}, {Stefanini}, {Tasso}, {Testi},
  {von der L{\"u}he}, {Valtier}, {Vannier}, \& {Ventura}}]{Millour2007}
{Millour}, F., {Petrov}, R.~G., {Chesneau}, O., {et~al.} 2007, \aap, 464, 107

\bibitem[{{Miroshnichenko} {et~al.}(2001){Miroshnichenko}, {Fabregat},
  {Bjorkman}, {Knauth}, {Morrison}, {Tarasov}, {Reig}, {Negueruela}, \&
  {Blay}}]{Miroshnichenko2001}
{Miroshnichenko}, A.~S., {Fabregat}, J., {Bjorkman}, K.~S., {et~al.} 2001,
  \aap, 377, 485

\bibitem[{{Monnier} {et~al.}(2002){Monnier}, {Tuthill}, \&
  {Danchi}}]{Monnier2002}
{Monnier}, J.~D., {Tuthill}, P.~G., \& {Danchi}, W.~C. 2002, \apjl, 567, L137

\bibitem[{{Monnier} {et~al.}(2007){Monnier}, {Tuthill}, {Danchi}, {Murphy}, \&
  {Harries}}]{Monnier2007}
{Monnier}, J.~D., {Tuthill}, P.~G., {Danchi}, W.~C., {Murphy}, N., \&
  {Harries}, T.~J. 2007, \apj, 655, 1033

\bibitem[{{Mourard} {et~al.}(2014){Mourard}, {Monnier}, {Meilland}, {Gies},
  {Millour}, {Benisty}, {Che}, {Ligi}, {Schaefer}, {Baron}, {Kraus}, {Zhao},
  {Pedretti}, {Berio}, {Clausse}, {Nardetto}, {Perraut}, {Spang1}, {Stee},
  {Tallon-Bosc}, {McAlister}, {ten Brummelaar}, {Ridgway1}, {Sturmann},
  {Sturmann}, {Turner}, \& {Farrington}}]{Mourard2014}
{Mourard}, D., {Monnier}, J.~D., {Meilland}, A., {et~al.} 2014, submited to
  \aap

\bibitem[{{Pourbaix}(1994)}]{Pourbaix1994}
{Pourbaix}, D. 1994, \aap, 290, 682

\bibitem[{{Pourbaix}(1998)}]{Pourbaix1998}
{Pourbaix}, D. 1998, \aaps, 131, 377

\bibitem[{{Quirrenbach} {et~al.}(1994{\natexlab{a}}){Quirrenbach}, {Buscher},
  {Mozurkewich}, {Hummel}, \& {Armstrong}}]{Quirrenbach_b1994}
{Quirrenbach}, A., {Buscher}, D.~F., {Mozurkewich}, D., {Hummel}, C.~A., \&
  {Armstrong}, J.~T. 1994{\natexlab{a}}, \aap, 283, L13

\bibitem[{{Quirrenbach} {et~al.}(1994{\natexlab{b}}){Quirrenbach},
  {Mozurkewich}, {Buscher}, {Hummel}, \& {Armstrong}}]{Quirrenbach_a1994}
{Quirrenbach}, A., {Mozurkewich}, D., {Buscher}, D.~F., {Hummel}, C.~A., \&
  {Armstrong}, J.~T. 1994{\natexlab{b}}, \aap, 286, 1019

\bibitem[{{Rosvick} \& {Scarfe}(1991)}]{Rosvick1991}
{Rosvick}, J.~M. \& {Scarfe}, C.~D. 1991, \mnras, 252, 68

\bibitem[{{Sana} {et~al.}(2012){Sana}, {de Mink}, {de Koter}, {Langer},
  {Evans}, {Gieles}, {Gosset}, {Izzard}, {Le Bouquin}, \&
  {Schneider}}]{Sana2012}
{Sana}, H., {de Mink}, S.~E., {de Koter}, A., {et~al.} 2012, Science, 337, 444

\bibitem[{{Schaefer} {et~al.}(2014){Schaefer}, {Brummelaar}, {Gies},
  {Farrington}, {Kloppenborg}, {Chesneau}, {Monnier}, {Ridgway}, {Scott},
  {Tallon-Bosc}, {McAlister}, {Boyajian}, {Maestro}, {Mourard}, {Meilland},
  {Nardetto}, {Stee}, {Sturmann}, {Vargas}, {Baron}, {Ireland}, {Baines},
  {Che}, {Jones}, {Richardson}, {Roettenbacher}, {Sturmann}, {Turner},
  {Tuthill}, {van Belle}, {von Braun}, {Zavala}, {Banerjee}, {Ashok}, {Joshi},
  {Becker}, \& {Muirhead}}]{Schaefer2014}
{Schaefer}, G.~H., {Brummelaar}, T.~T., {Gies}, D.~R., {et~al.} 2014, \nat,
  515, 234

\bibitem[{{Shao} \& {Colavita}(1992)}]{Shao1992}
{Shao}, M. \& {Colavita}, M.~M. 1992, \aap, 262, 353

\bibitem[{{Siess} {et~al.}(2013){Siess}, {Izzard}, {Davis}, \&
  {Deschamps}}]{Siess2013}
{Siess}, L., {Izzard}, R.~G., {Davis}, P.~J., \& {Deschamps}, R. 2013, \aap,
  550, A100

\bibitem[{{Tango} {et~al.}(2009){Tango}, {Davis}, {Jacob}, {Mendez}, {North},
  {O'Byrne}, {Seneta}, \& {Tuthill}}]{Tango2009}
{Tango}, W.~J., {Davis}, J., {Jacob}, A.~P., {et~al.} 2009, \mnras, 396, 842

\bibitem[{{Tycner} {et~al.}(2011){Tycner}, {Ames}, {Zavala}, {Hummel},
  {Benson}, \& {Hutter}}]{Tycner2011}
{Tycner}, C., {Ames}, A., {Zavala}, R.~T., {et~al.} 2011, \apjl, 729, L5

\bibitem[{{Zhao} {et~al.}(2008){Zhao}, {Gies}, {Monnier}, {Thureau},
  {Pedretti}, {Baron}, {Merand}, {ten Brummelaar}, {McAlister}, {Ridgway},
  {Turner}, {Sturmann}, {Sturmann}, {Farrington}, \& {Goldfinger}}]{Zhao2008}
{Zhao}, M., {Gies}, D., {Monnier}, J.~D., {et~al.} 2008, \apjl, 684, L95

\end{thebibliography}

\end{document}